\documentstyle[12pt,epsf]{article}

\newcommand{\be}{\begin{equation}}
\newcommand{\ee}{\end{equation}}
\newcommand{\ben}{\begin{eqnarray}\displaystyle}
\newcommand{\een}{\end{eqnarray}}
\newcommand{\refb}[1]{(\ref{#1})} 
\newcommand{\p}{\partial}
\newcommand{\sectiono}[1]{\section{#1}\setcounter{equation}{0}}

\newcommand{\rr}{R}
\newcommand{\II}{{\cal I}}

\newcommand{\SS}{{\cal S}}

\newcommand{\HH}{{\cal H}}

\newcommand{\OO}{{\cal O}}

\newcommand{\RR}{{\cal R}}

\newcommand{\QQ}{{\cal Q}}
  
\newcommand{\bsliv}{\langle\Xi |}

\newcommand{\wt}{\widetilde}
\newcommand{\wh}{\widehat}

\newcommand{\del}{\Delta}
\newcommand{\vp}{\varphi}

\begin{document}
{}~
\hfill\vbox{\hbox{hep-th/0105168}\hbox{CTP-MIT-3131}
\hbox{PUPT-1986} \hbox{NSF-ITP-01-34}
}\break

\vskip .6cm

\centerline{\large \bf 
Boundary CFT Construction of D-branes in}

\medskip

\centerline{\large \bf Vacuum String Field Theory }

\vspace*{4.0ex}

\centerline{\large \rm Leonardo Rastelli$^a$, Ashoke Sen$^b$ and Barton
Zwiebach$^c$}

\vspace*{4.0ex}

\centerline{\large \it ~$^a$Department of Physics }

\centerline{\large \it Princeton University, Princeton, NJ 08540, USA}

\centerline{E-mail:
        rastelli@feynman.princeton.edu}

\vspace*{2ex}

\centerline{\large \it ~$^b$Harish-Chandra Research
Institute}

\centerline{\large \it  Chhatnag Road, Jhusi,
Allahabad 211019, INDIA}

\centerline {and}
\centerline{\large \it Institute for Theoretical Physics}

\centerline{\large \it 
University of California, Santa Barbara, CA 93106, USA
}

\centerline{E-mail: asen@thwgs.cern.ch, sen@mri.ernet.in}

\vspace*{2ex}

\centerline{\large \it $^c$Center for Theoretical Physics}

\centerline{\large \it
Massachussetts Institute of Technology,}

\centerline{\large \it Cambridge,
MA 02139, USA}

\centerline{E-mail: zwiebach@mitlns.mit.edu}

\vspace*{5.0ex}

\centerline{\bf Abstract}
\bigskip

In previous papers we built (multiple) D-branes  
in flat space-time as classical solutions of 
the string field theory based on the tachyon vacuum. 
In this paper we construct classical solutions 
describing all D-branes in  any fixed space-time
background defined by a two dimensional CFT  
of central charge 26.  
A key role is played by the geometrical definition 
of the sliver state in general boundary CFT's. 
The correct values for ratios of  D-brane tensions 
arise because the norm of the sliver solution is naturally 
related to the disk partition function of the
appropriate boundary CFT.
We also explore the possibility of reproducing the known   
spectrum of physical states on a D-brane as deformations 
of the sliver.

\vfill \eject

\baselineskip=16pt

\tableofcontents

\newpage

\sectiono{Introduction and summary}  \label{s1} 

Much work has been done in understanding 
various conjectures about tachyon
condensation on D-branes in bosonic string
theory \cite{9902105,9904207,9911116} using 
cubic open string field
theory \cite{WITTENBSFT} (SFT).
Although
the results are very 
impressive, they ultimately rely 
on numerical study of the solutions
of the equations of motion using the level truncation  
scheme~\cite{KS,9912249,0002237,0002117,0003031,
0005036,0006240,0007153,
0008033, 0008053,0008101,0009105,0010190,
0011238,0101014,0101162,0103085,0103103,
0102085,0105024}.\footnote{For some early
attempts at understanding the open string  tachyon, 
see refs.\cite{EARLY}.
For field theory models of tachyon condensation, see 
refs.\cite{FIELD1,FIELD2}. Study of tachyon condensation using 
renormalization  group approach has been carried 
out in refs.\cite{RG}.} 
In 
a series of three 
papers~\cite{0012251,0102112, TWO} we 
attempted an analytic
approach to the issues of tachyon condensation  
by proposing a candidate string 
field theory action which describes string
field theory expanded around the tachyon vacuum. 
As opposed to the conventional cubic SFT where the kinetic operator
is the BRST operator $Q_B$, here the kinetic operator $\QQ$ is
non-dynamical
and is built solely out of worldsheet ghost
fields.\footnote{A subset
of this class of actions was discussed previously 
in ref.~\cite{HORO}.}
In this class of 
actions
the absence of physical open string states around the vacuum is
manifest.
Gauge invariance holds, 
and therefore basic
consistency requirements are expected to be satisfied.
Furthermore we showed that this theory 
contains classical solutions
representing D-$p$-branes for all $p\le 25$, with correct ratios of
tensions, thereby providing a non-trivial check on the correctness of 
our proposal. 
The key ansatz that made this analysis possible was that the
string field solution 
representing a D-brane factorizes into a ghost part
$\Psi_g$ and a matter part $\Psi_m$, with $\Psi_g$ 
the same for
all D-branes, and $\Psi_m$ 
different for different D-branes.
The $\Psi_m$ representing the D25 brane is a particular surface
state of  the corresponding boundary conformal field theory (BCFT) known 
as the sliver~\cite{0006240, 0008252, 0102112,TWO}.

In this paper we discuss the construction of D-brane solutions
in this theory using  conformal field theory techniques.
This construction has certain conceptual and practical advantages. In
the 
construction of ref.~\cite{0102112} we obtained an expression for the
ratio of tensions of different D-branes in 
terms of ratios of determinants
of infinite matrices. There was no analytic 
understanding, however,  why this
ratio gave the expected answer.  In the approach taken here, 
the correct values
of all ratios of 
tensions are obtained  manifestly. Furthermore the
analysis given in this paper can be carried 
out in the background of any bulk
conformal field theory (CFT), for 
any D-brane  
described by an appropriate 
BCFT.\footnote{Throughout this paper 
a specific BCFT will refer to the boundary conformal 
field theory associated with a {\it single} D-brane. Multiple D-brane 
solutions will be obtained by first constructing solutions corresponding 
to the individual D-branes, and then taking appropriate superposition of 
the solutions.}
                                                               
We start with some fixed space-time background described by a particular 
bulk CFT, and choose once and for all a specific reference BCFT, 
denoted as  BCFT$_0$. 
We define our string field to be a state in the Hilbert space of
BCFT$_0$. 
In the analysis of refs.\cite{0012251,0102112,TWO} the D25-brane BCFT 
played the role of BCFT$_0$.
As in those works, we seek solutions of 
the form $\Psi_g\otimes \Psi_m$, where the ghost component 
$\Psi_g$ is universal, but
the matter part $\Psi_m$ varies from one solution to another.
Under this factorization hypothesis, the matter
part of the string field satisfies 
a simple equation: it squares to itself
under $*$-multiplication. 
One particular solution of this equation is the matter part of the
sliver 
state of BCFT$_0$. We identify 
this state as the solution representing the
D-brane associated with BCFT$_0$, generalizing  the identification of
the 
D25-brane solution as the sliver of the corresponding BCFT
\cite{0102112,TWO}.

In the conformal field theory
description~\cite{0006240},   the sliver is regarded as a
surface state by using the standard 
procedure for associating a BCFT state 
to  every Riemann surface with a boundary, with
one puncture at the boundary and a
local coordinate at the puncture. 
The sliver is described as the 
surface state associated 
with a specific  
once-punctured disk.
This description of the sliver is universal in the sense that
the state takes
exactly the same form for any BCFT when written
in terms of the Virasoro operators of the BCFT. 
It is therefore  natural to expect  
that, just as in the case of BCFT$_0$, the sliver of an arbitrary BCFT 
describes the D-brane  associated with that BCFT. 
In order to regard different D-branes as different solutions 
in the SFT, however, we need to express the slivers associated with these
different  BCFT's as states in the Hilbert space of BCFT$_0$.
We find an explicit algorithm for doing this.
The solution
constructed this way correctly reproduces the tension of the D-brane
associated with the specific BCFT (up to an overall normalization  
constant 
which is the same for all D-branes). This is a
nontrivial result  that
follows because the tension, in vacuum SFT, is given by the BPZ
inner product of the solution with itself,
and we show that this inner product is simply related to the disk
partition function in the specific BCFT. 
The identification of 
the disk partition function with the D-brane tension is 
a well-known result~\cite{9511173,9707068, 9807161,9909072,0101200}  
that has played
a crucial role in the study of tachyon
condensation in boundary
string field theory~\cite{9208027,9210065,9303067,9303143,
9311177,0008231,
0009103,0009148,0009191,0011009}.

\medskip
The paper is organized as follows. In section \ref{sr1} we give a
brief summary 
of the results of refs.\cite{0012251,0102112}. In section 
\ref{s2} we give a detailed
review of the construction of the sliver as a surface state
associated with a once-punctured disk with a local coordinate at the
puncture. We describe it in various coordinate
systems which are
useful in our analysis. We also explain why 
it is a well defined state, in
that the inner product of the sliver with any BCFT 
state 
associated with a vertex operator is computable and finite.
This is not a priori obvious, since the state arises in a limit where the
local coordinate at the puncture becomes singular. Nevertheless,
SL(2,R) invariances of the state allow geometrical representations
where the local coordinate is non-singular, and the resulting
state is manifestly regular.  
It is significant that even the regular geometrical description
of the sliver is unusual in that the open string midpoint, which
usually is somewhere inside the disk, 
reaches the boundary of the disk. This fact gives 
an intuitive understanding
of the left-right factorization of the sliver functional, a 
key result in the analysis 
of \cite{TWO,gross-taylor}.  Our 
discussion also addresses star
multiplication of surface states, and explains 
why the sliver squares to itself under star multiplication.

In section \ref{s3} we use the  universal
description of the
sliver to construct 
solutions of the SFT equations of motion describing different
D-branes  in some fixed space-time background.
We take for the space-time background some fixed bulk CFT with central 
charge 26, and choose a reference 
boundary conformal field theory BCFT$_0$, in whose 
state space the string field takes value.  
The matter part of the sliver of BCFT$_0$  
is the matter part of a 
solution 
describing the D-brane 
corresponding to BCFT$_0$. We show
in section \ref{sn1} 
 that every other
D-brane in this  space-time 
background, described 
by some other boundary conformal field theory BCFT$'$, can also be 
obtained 
as a solution in this SFT.
We explicitly write down the classical solution 
describing this D-brane, and verify that it satisfies the equations of 
motion and correctly reproduces the tension of the D-brane.
In section \ref {snew4.2} we show that 
solutions describing D-branes associated with two {\it different}
BCFT$'$s $*$-multiply to zero, and hence we can construct multiple
D-brane solutions by superposing them. 
We also give a  construction of  identical
coincident D-branes.

In section \ref{s4a} we discuss the construction of classical solutions 
corresponding to a two dimensional field theory obtained by deforming
BCFT$_0$ by a  relevant or marginal boundary operator. The results bear 
strong resemblance to those of boundary string field theory, except for
one important  difference. In boundary string field theory, the
coefficient of a relevant deformation is driven to infinity 
(or more generally to the infrared fixed point)  
by the equations of motion. 
In contrast, here we get a solution of the
equations of motion for arbitrary value of this coefficient. 
When we
compute the tension of the corresponding solution in section \ref{sn0},
however, 
we recover the
partition function on a  disk of the deformed boundary {\em conformal} 
field theory (which we  call BCFT$'$), with the
coefficient of
the perturbation driven to 
its infrared fixed point 
due to a conformal transformation
involving infinite rescaling. Thus the conclusion 
is that different values
of these coefficients describe the same D-brane solution
$-$ the one associated with BCFT$'$.\footnote{This
had been earlier anticipated by Witten~\cite{PRIVATE}.} For 
exactly marginal 
deformations, of course, different values of the coefficient represent
genuinely different solutions.

If BCFT$'$ and BCFT$_0$ are related by an 
exactly marginal deformation, we can also study small
deformations of the solution by taking the perturbing parameter to be
small. This is done in section \ref{s4.2}. We show in section \ref{ss31} 
that these small deformations can be thought of as 
covariant derivatives of the sliver state 
with respect to
a canonical  theory-space 
connection introduced in ref.~\cite{RSZ}. 
We also  discuss the background independence of vacuum string field
theory in the language of connections over theory space.

In section \ref{s6} we address the question of the spectrum of physical
states around the 
solution describing the D-brane associated with BCFT$_0$. 
Since
general excitations around this solution involve both matter and ghost
oscillators, this problem cannot be studied 
completely without knowing the
form
of the kinetic operator $\QQ$ and the ghost part 
$\Psi_g $ of the
D-brane solution. We make an attempt to study this problem in section
\ref{s4.1} by restricting
ourselves to excitations of the factorized type with the same universal
ghost part $\Psi_g $ but arbitrary fluctuations in the matter
sector. We 
find that we do get solutions of the linearized equations of
motion for  every dimension-one primary in the matter
part of BCFT$_0$, as is  
expected of
the open string spectrum on a D-brane.
We also need to determine which of
these solutions could be related by linearized gauge transformations.
Again, 
without detailed knowledge of $\QQ$, 
we proceed with a restricted
class of gauge transformation parameters described in section
\ref{ssgauge} which preserve the factorized form of the linearized
fluctuations. Section
\ref{s6.1.1} describes some subtleties  
that arise in studying 
the normalization of states. 
We discuss a plausible 
resolution 
leading to some
interesting conclusions. First, it is strongly suggested that
the kinetic operator $\QQ$ of vacuum SFT must annihilate the
identity operator. Second, we find that any linearized
solution to the equations of motion with finite BPZ norm
is pure gauge. Physical states are then argued to be associated
with states whose BPZ norm diverges logarithmically in the
parameters defining finite versions of the sliver. 
Although these criteria lead us to identify sliver 
deformations by dimension-one primaries as physical states,
it is not clear how dimension one nulls and
non-primary operators are removed from the spectrum.

We conclude 
in section \ref{s5} with a discussion of
some general aspects of the string field theory around the tachyon
vacuum.  

\sectiono{Review of vacuum string field theory} 
\label{sr1}

In this section we shall briefly describe the results of
refs.\cite{0012251,0102112}.
In these papers we proposed a
form of
the string field theory action around the open bosonic string
tachyon vacuum
and discussed classical solutions
describing D-branes of various dimensions.
In order to write concretely this theory we choose
to use the state space $\HH$ of the combined matter-ghost boundary
conformal field theory (BCFT) describing the D25-brane.
The string
field  
$\Psi$ is a state of ghost number one  
in $\HH$  and the
string field action is given by: 
\be \label{eo1}
\SS (\Psi) \equiv \,-\, {1\over g_0^2}\,\,\bigg[\, {1\over 2} \langle
\,\Psi \, ,
 \, \QQ\, \Psi
\rangle + {1\over 3}\langle \,\Psi \, , \, \Psi *
\Psi \rangle \bigg] \,,
\ee
where $g_0$ is the open
string coupling constant, $\QQ$ is an operator made purely of ghost
fields, $\langle \, ,  \, \rangle$
denotes the BPZ inner product,
and $*$ denotes the usual $*$-product of the string
fields~\cite{WITTENBSFT}. 
$\QQ$ satisfies the requirements:
\ben \label{eFINp}
&& \QQ^2 = 0, \nonumber \\
&& \QQ (A * B) = (\QQ A) * B + (-1)^{A} A * (\QQ B)\, , \\
&& \langle \, \QQ A , B \,\rangle = - (-)^A \langle A , \QQ B \rangle
\,. \nonumber
\een 
The action \refb{eo1} is then invariant under the gauge transformation:
\be \label{egtrs}
\delta\Psi = \QQ\Lambda + \Psi * \Lambda - \Lambda * \Psi \, ,
\ee
for any ghost number zero state $\Lambda$ in $\HH$. Ref.\cite{0012251}
contains candidate operators $\QQ$ satisfying these constraints; for our
analysis we shall not need to make a specific choice of $\QQ$.
The equations of motion are
\be \label{eo2}
\QQ \Psi + \Psi * \Psi = 0\, .
\ee

In ref.~\cite{0102112} we made the ansatz that all D-$p$-brane
solutions in this theory have the factorized form:
\be \label{eo3}
\Psi = \Psi_g \otimes \Psi_m\, ,
\ee
where $\Psi_g$
denotes a state obtained by acting with the ghost
oscillators on the SL(2,R) invariant vacuum of the ghost BCFT, and
$\Psi_m$  is a
state obtained by acting with matter operators on the SL(2,R)
invariant
vacuum of the matter BCFT.
Let us denote by
$*^g$ and $*^m$ the star product in the ghost and matter sector
respectively.
Eq.\refb{eo2} then factorizes as
\be \label{eo4}
\QQ \Psi_g = - \Psi_g *^g \Psi_g \,,
\ee
and
\be \label{eo5}
\Psi_m = \Psi_m *^m \Psi_m\, .
\ee
We further assumed that the ghost part $\Psi_g$ is universal for all
D-$p$-brane solutions. Under this assumption the ratio of energies
associated with two different D-brane solutions, with matter parts
$\Psi_m'$ and $\Psi_m$ respectively, is given by:
\be \label{eo7}
{\langle \Psi_m' | \Psi_m'\rangle_m \over \langle \Psi_m |
\Psi_m\rangle_m} \, ,
\ee
with $\langle \cdot| \cdot\rangle_m$ denoting BPZ inner product in
the matter BCFT. Thus the ghost part drops out of this calculation.

In ref.~\cite{0102112} we constructed analytically 
the matter part of the solution for
different D-$p$-branes, and verified numerically 
that we get the correct ratio of
tensions of D-$p$-branes using eq.\refb{eo7}. The matter part of the
D-25-brane solution was given by the sliver 
state $|\Xi\rangle$ which will
play an important role in the analysis of this paper. The construction of
$|\Xi\rangle$ will be reviewed in section \ref{s2}.

In this paper, we shall be using a more general setup. Instead of 
restricting ourselves to D-branes in flat space-time, we shall consider
a 
general space-time background described by some arbitrary bulk CFT. The 
role of the D25-brane, in whose Hilbert space the string field takes 
value, is played by some fixed D-brane in this background associated
with 
a specific BCFT. We shall call this BCFT$_0$. The matter part of 
$|\Xi\rangle$, described as a surface state in BCFT$_0$, then describes 
the 
D-brane associated with BCFT$_0$.

\sectiono{The various pictures of the sliver} \label{s2}

\newcommand{\lo}{\xi}
\newcommand{\gl}{z}
\newcommand{\gf}{\wh w}

In this section we will examine the sliver state $|\Xi\rangle$
from its geometrical definition. In this context we take the
opportunity to describe in detail the various ways
the sliver can be presented.  
We emphasize how, despite
its origin as a surface state with a singular coordinate, it
is a well-defined state satisfying $\Xi*\Xi=\Xi$. 
Although we are interested in the matter part of the sliver, we shall
work in most part with the full sliver
including the
ghost sector, so that there is no central charge contribution to the
various conformal transformations and gluing operations. Having
established  that $\Xi*\Xi=\Xi$, we can then use the factorization
property of the $*$-product to conclude that $\Xi_m*^m\Xi_m=\Xi_m$ with
suitable normalization of $\Xi_m$ 
(which could be infinite, but is universal in the sense that it does not 
depend on the specific choice of matter BCFT).

\subsection{Viewpoints on surface states} \label{sq0}

The sliver is a ghost number zero state that has a 
universal 
definition. It is a {\it surface state},
which means that for any given BCFT 
it can be defined as the bra $\langle \Xi|$
associated to a particular Riemann surface $\Sigma$.  
The surface in question
is a disk $D$ with one puncture $P$ at the boundary. 
Moreover, there
is a local coordinate at this puncture.
The local coordinate at the puncture for the case of the sliver
is obtained by a limiting procedure, to be reviewed and
elaborated below. 

\begin{figure}[!ht]
\leavevmode
\begin{center}
\epsfxsize = 15 cm 
\epsfbox{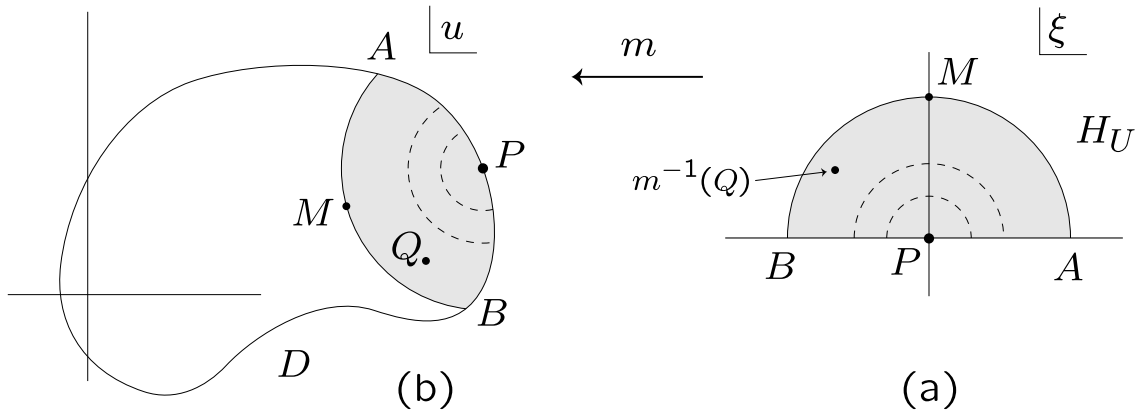}
\end{center}
\caption[]{\small   A punctured disk $D$ with a local coordinate
around the puncture $P$. The coordinate is defined through a map
$m$ from a canonical half disk $H_U$ to the disk. The arcs
$AM$ and $MB$ in $D$ represent the left half and the right
half of the open string respectively.} \label{f1}
\end{figure}

We shall begin with a general discussion of
surface states associated with a disk with one puncture. 
A local coordinate at a puncture is obtained from 
an analytic map $m$ taking a canonical half-disk $H_U$
defined as 
\be
\label{halfdisk}
H_U: \,\,\{ |\lo|\leq 1, \Im (\lo)\geq 0\}\,,
\ee
{\it into} $D$, where $\lo=0$ maps to the puncture $P$, and
the image of the real segment
$\{|\lo|\leq 1, \Im (\lo) =0\}$ lies on the boundary of $D$. 
The coordinate $\lo$ of the half disk is
called the {\it local coordinate}. For any point $Q\in D$ in the
image of the map, 
$\lo( m^{-1}(Q))$ 
is the local coordinate of the point.
Using any {\it global  
coordinate} $u$ on the disk $D$, 
the map $m$ can be
described by some analytic function $s$: 
\be u = s(\lo)\,, \quad  u (P)= s(0) \,.
\ee
Figure \ref{f1} shows a disk $D$ with a local coordinate around
the puncture $P$. 
The image under $m$
of the semicircle 
$|\lo| =1$, corresponding to the curve $AMB$ in Fig.\ref{f1}  is usually
referred to as the open string. 
The image of point $M$,
corresponding to $\lo=i$, is called the string midpoint.
The image of the arc $AM$
is called the left-half 
(as seen from the interior of $D$)
of the open string, and similarly
the image
of $MB$ is called the right-half 
of the open string. 
If we denote by $\sigma={1\over i} \ln\lo$ the coordinate along the
string,  then the left-half of the string corresponds to the region 
$0\le\sigma<{\pi\over 2}$ and the right-half of the string corresponds
to 
the region ${\pi\over 2}<\sigma\le\pi$.
The image of the
half-disk $H_U$ in the $u$-plane, shown by the shaded region in
Fig.\ref{f1}, will be called the {\it local coordinate patch.}

Given this geometrical data, and a BCFT with state space
${\cal H}$, the state 
$\langle\Sigma|\in {\cal H}^* $
associated to the surface $\Sigma$ is defined as follows.
For any local operator $\phi (\lo)$, with associated state
$|\phi\rangle = \lim_{\lo\to 0} \phi(\lo) |0\rangle$ we set
\be \label{esdef}
\langle\Sigma| \phi \rangle = \langle  s
\circ \phi (0) \rangle_{D} \,,
\ee
where $\langle ~\rangle_{D}$ 
corresponds to correlation function on $D$
and $s\circ \phi (0)$ 
denotes the conformal
transform of the operator by the map $s(\xi)$, {\it i.e.} the operator
$\phi(\lo=0)$
expressed using the appropriate conformal map in terms of
$\phi(s(0))$. 
For a
primary of dimension
$h$, $s\circ \phi (0) = \phi(s(0))
(s'(0))^h$. The right
hand side of eq.\refb{esdef} can be interpreted as
the one point
function on $D$ of the local operator $\phi$ inserted at $P$ using the
local coordinate
$\lo$ defined there. We also call, with a small abuse of notation,
$|\Sigma\rangle \in \HH$ a surface state; this is simply
the BPZ conjugate 
of $\langle \Sigma|$.  
While computations of correlation functions
involving states in $\HH$  requires that the map $s$ be defined
only locally around the puncture $P$, more general constructions, such as
the gluing of surfaces, an essential tool in the operator formulation of
CFT, requires that the full map of the half disk $H_U$ into
the disk $D$  be well
defined.

At an intuitive level
$\langle\Sigma|$ can  
be given the following functional integral representation. Consider the
path integral over the basic elementary fields of the two dimensional
conformal field theory, $-$ collectively denoted as $\vp$, $-$ on the
disk $D$ minus the local coordinate patch, with some fixed boundary
condition $\vp=\vp_0(\sigma)$ on the boundary $AMB$ of the local
coordinate patch, and the open string boundary condition corresponding to
the BCFT under study on the rest of the boundary of this region. The
parameter $\sigma$ is the coordinate labeling the open string along
$AMB$, defined  
through $\xi=e^{i\sigma}$. 
The result of 
this path integral
will a functional of the boundary value $\vp_0(\sigma)$. We
identify this as the
wave-functional of the state $\langle\Sigma|$. (For describing the
wave-functional of $|\Sigma\rangle$ we need to make a
$\sigma\to(\pi-\sigma)$ transformation.) 
On the other hand the wave-functional
of the state $|\phi\rangle$ can be
obtained by performing the path integral over $\vp$ on the unit
half-disk in the $\xi$ coordinate system, with the boundary condition
$\vp=\vp_0(\sigma)$ on the
semicircle, open
string boundary condition corresponding to the BCFT
on real axis, and a vertex operator $\phi(0)$ inserted at the origin.
We can now compute
$\langle\Sigma|\phi\rangle$ for any state $|\phi\rangle$ in $\HH$ by
multiplying the two wave-functionals and integrating over
the argument 
$\vp_0(\sigma)$. The net result is a path integration
over $\vp$ on the full disk $D$, with the boundary condition
corresponding to BCFT over the full boundary and a vertex operator
$\phi$ inserted at the puncture $P$ {\it using the $\xi$ coordinate
system.} This is precisely eq.\refb{esdef}.

\newcommand{\ff}{\check f}

\begin{figure}[!ht]
\leavevmode
\begin{center}
\epsfxsize = 15 cm \epsfbox{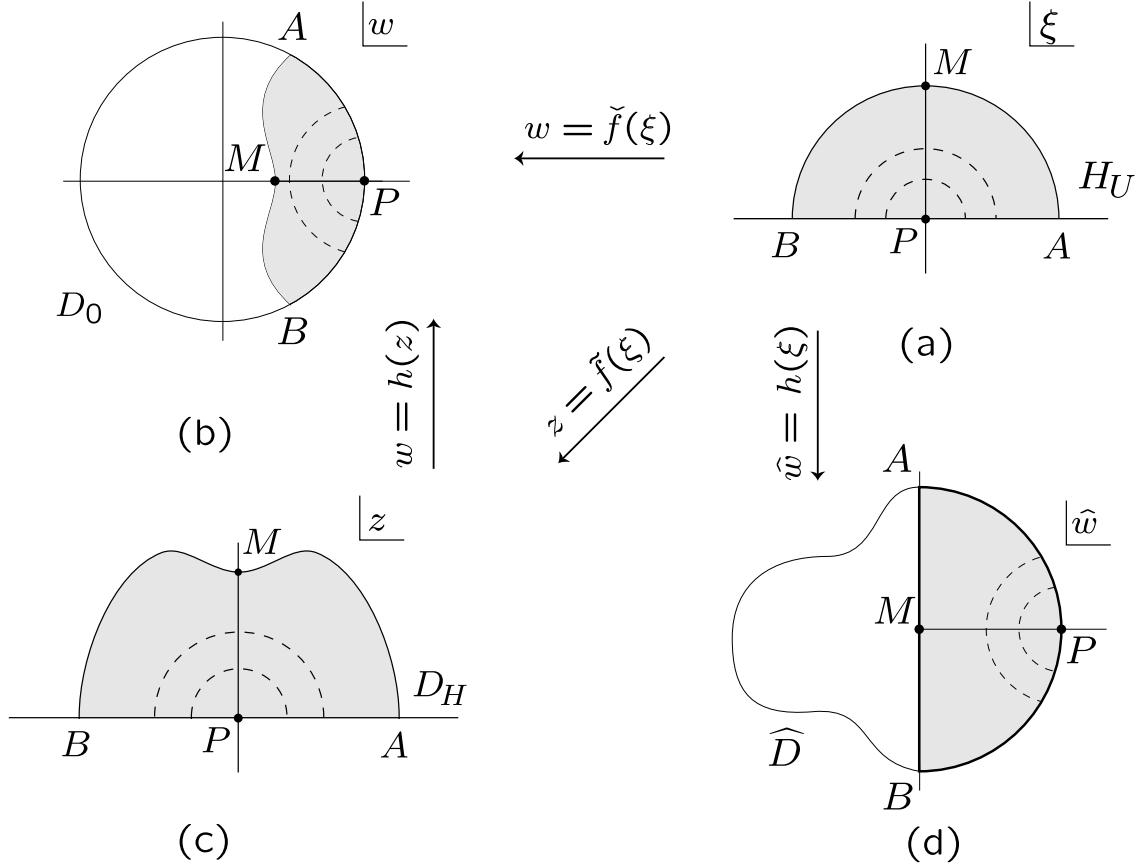 }
\end{center}
\caption[]{\small Three canonical presentations of the disk $D$.
(b) Disk presented as the unit disk $D_0$ with global coordinate $w$.
(c) Disk presented as the upper half plane $D_H$ with global coordinate
$z$. (d) Disk presented as $\wh D$ with global coordinate
$\wh w$. Here the image of $H_U$ is also a half-disk.
} \label{f2}
\end{figure}

For future use, we shall now describe three 
canonical ways of defining the surface state
$\langle\Sigma|$, using three  
presentations of the disk $D$.
In the first one, we present $D$ as the  unit disk $D_0: |w|\leq 1$
in a $w$-plane. The puncture will be located at $w=1$, and the local
coordinate described as
\be
\label{wcoor}
w = \ff(\xi)\,, \quad w(P) = \ff(0) = 1\,.
\ee 
This is shown in Figure \ref{f2}(b). In this presentation we have
that eq.\refb{esdef} takes the form
\be \label{stw}
\langle\Sigma| \phi \rangle = \langle   \ff
\circ \phi (0) \rangle_{D_0}
\ee
In the second presentation of $D$ we map it to the
upper half plane (UHP) with global coordinate $\gl$,
and locate the puncture at $\gl=0$. We will denote
this upper half-plane as $D_H$.
More concretely, we define $\gl$ through the   relation 
\be \label{ed1}
w = h(\gl) \equiv  {1+ i\gl\over 1-i\gl}\, ,
\ee
which maps the UHP, labeled by $\gl$, 
to the unit disk $D_0$. We then have that
the local coordinate around $P$ in this presentation takes 
the form: 
\be \label{ed2}
z = h^{-1}(w) = h^{-1} (\ff(\lo)) \equiv  \wt f(\lo) 
\ee
In this presentation eq. \refb{esdef} can be rewritten as 
\be \label{ed3}
\langle\Sigma| \phi \rangle = 
\langle  \wt f\circ \phi (0) \rangle_{D_H}\, ,
\ee
where $\langle ~\rangle_{D_H}$ denotes correlation function on the upper
half plane. The disk $D_H$ with its coordinate is shown in 
figure \ref{f2}(c). 

Finally, we introduce the third presentation, where the disk $D$
is mapped into a disk $\wh D$ having the special property that
the local coordinate patch, {\it i.e.} the 
image of the half disk $H_U$ in
$\wh D$, is particularly simple. With
global coordinate $\wh w$, 
the image of $H_U$ appears  as a vertical
half-disk of unit radius,
with the curved part of $H_U$ mapped to
the imaginary axis and the diameter 
of $H_U$ mapped to the unit semi-circle
to the right of the imaginary axis
(see Fig.\ref{f2}(d)). This is achieved by taking,  for
$\xi\in H_U$, 
\be \label{ed4}
\gf = h(\xi) = {1 + i\xi \over 1 - i\xi}\, .
\ee
In this presentation the rest of $\wh D$ 
may take a complicated form. 
We can now rewrite eq.\refb{esdef} as 
\be \label{ed5}
\langle\Sigma| \phi \rangle = 
\langle  h\circ \phi (0) \rangle_{\wh D}\,,
\ee
where $\langle~ \rangle_{\wh D}$ denotes the correlation function on the
disk $\wh D$ with appropriate open string boundary condition 
at the
boundary
of $\wh D$. In this description the information about the state is
encoded in the shape of the disk $\wh D$.

\subsection{The sliver surface state defined} \label{sq1}

\begin{figure}[!ht]
\leavevmode
\begin{center}
\epsfxsize = 15 cm \epsfbox{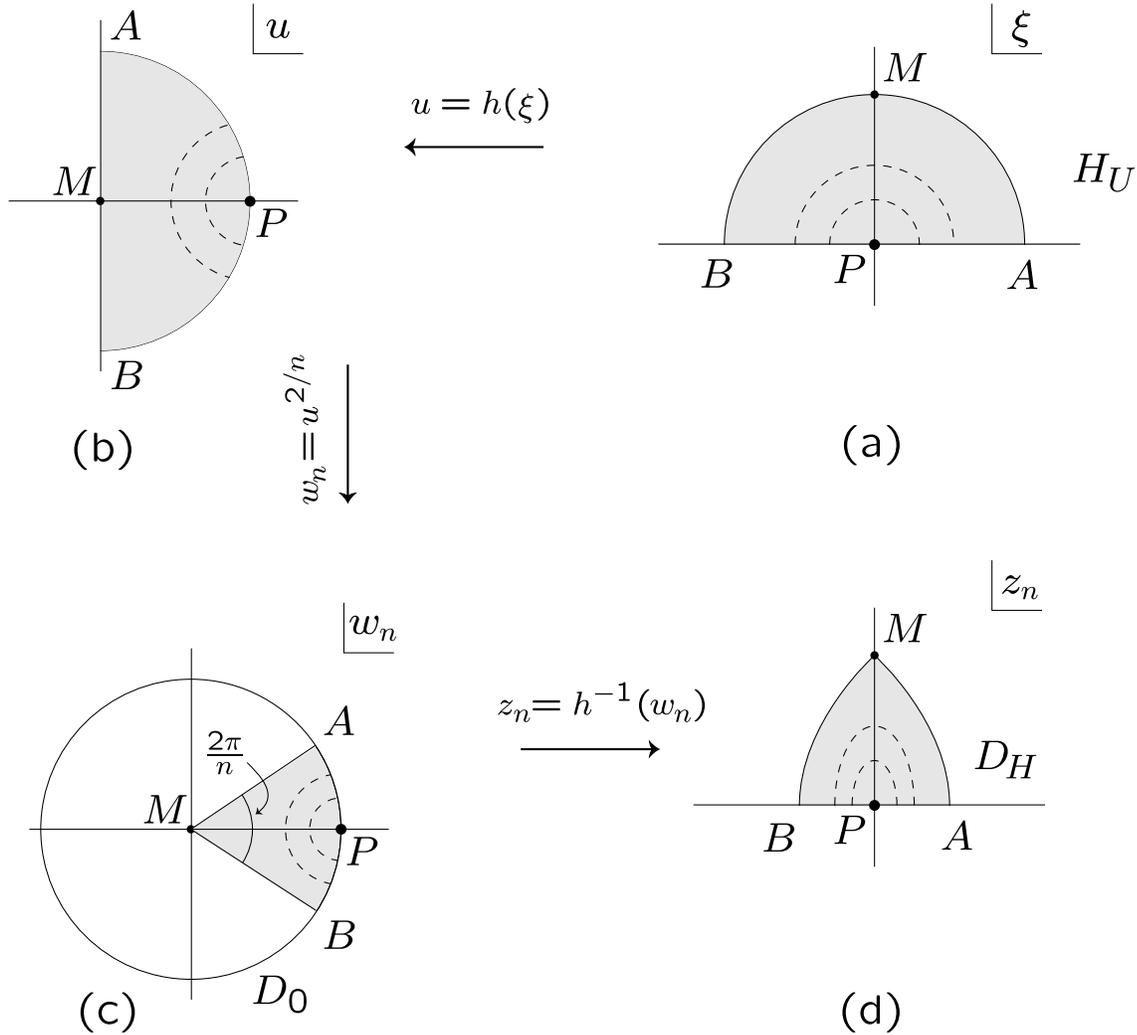 }
\end{center}
\caption[]{\small The definition of wedge states for arbitrary
$n$. (a) The canonical half disk $H_U$. (b) The map of
$H_U$ into a vertical half-disk. (c) The inclusion of the
vertical half-disk of (b) into the disk $D_0$ with global
coordinate $w$. (d) The map of $D_0$ into the upper half
plane $D_H$.
} \label{f3}
\end{figure}

\begin{figure}[!ht]
\leavevmode
\begin{center}
\epsfxsize = 15 cm \epsfbox{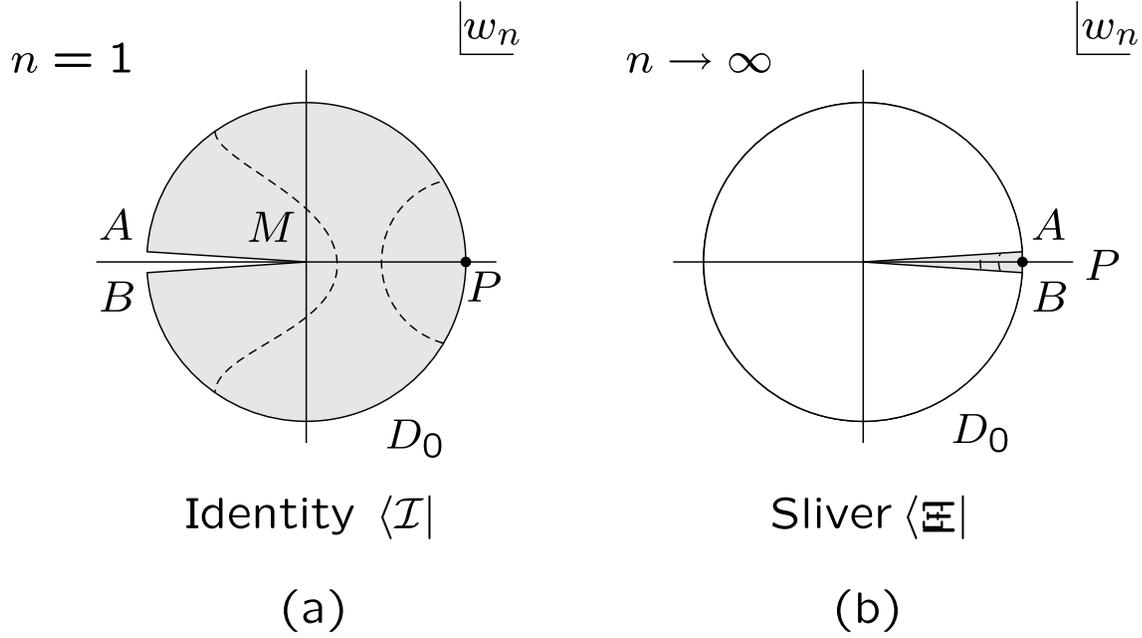 }
\end{center}
\caption[]{\small (a) The surface state corresponding to
the identity string field $\langle\II|$. Here the image
of $H_U$ covers the full disk, except for a cut in the
negative real axis. (b) The surfaces state corresponding
to the sliver $\langle \Xi|$. Here the image of $H_U$ covers
an infinitesimally thin sliver around the positive real axis. 
} \label{f4}
\end{figure}

\begin{figure}[!ht]
\leavevmode
\begin{center}
\epsfxsize = 15 cm \epsfbox{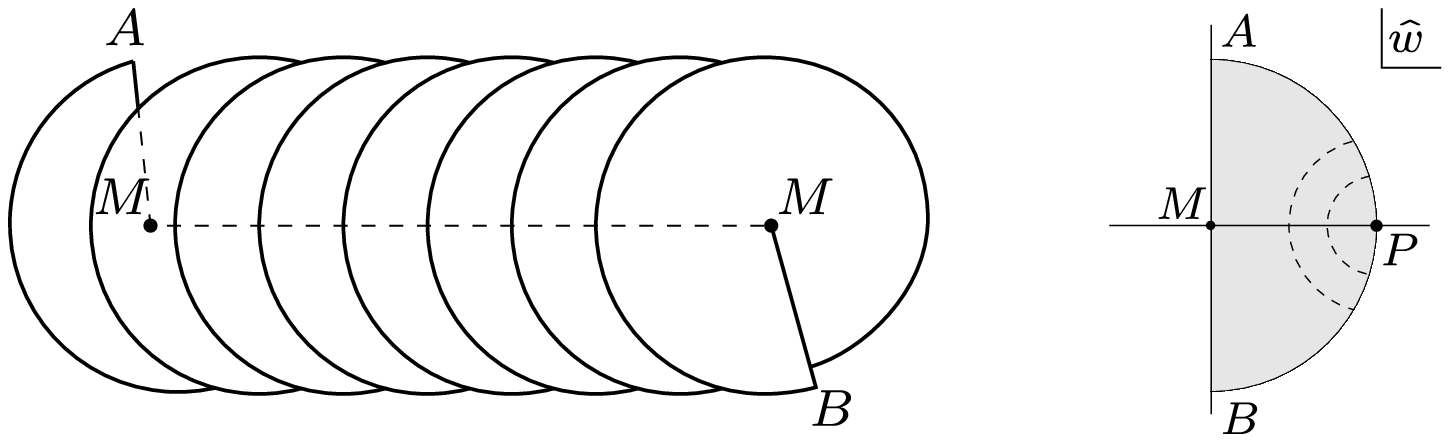 }
\end{center}
\caption[]{\small Using $\wh w$ global coordinates the
sliver appears as a cone with infinite excess angle-- namely,
an infinite helix. The segments $AM$ and $BM$ represent the
left-half and the right-half of the string.  The local 
coordinate patch, 
represented by the shaded half disk shown to the right, must be
glued in to form the complete surface.
} \label{f9}
\end{figure}

We can now define the sliver state following 
the route originally taken through a limit of 
certain `wedge states' \cite{0006240}. 
We shall give this description in all three pictures by explicitly
specifying the maps $\ff(\lo)$, $\wt f(\lo)$ and the disk $\wh D$.
In doing so we will refer to Fig.\ref{f3}, \ref{f4}, and \ref{f9}.
We begin by giving
the description on the unit disk
$D_0$. We define for any positive real number 
$n >  0$   
\be
\label{ih}
w_n = \ff_n (\xi) \equiv (h(\lo))^{2/n} =  
\Big({1+ i\lo\over 1-i\lo}\Big)^{2/n}\,, 
\ee
which for later purposes we also write as
\be
\label{ef1}
w_n = \exp \Bigl( \, i \,{4\over n} \, \tan^{-1} (\lo) \Bigr) \,.
\ee
As already pointed out before,
the map $h(\lo)$ takes the canonical half disk into
a similar presentation, with the puncture now on the
curved side of a half-disk (Figure \ref{f3}-b). Moreover the
string midpoint $M$ at $\lo=i$  is mapped to $h(i)=0$. The map
$w_n = (h(\lo))^{2/n}$
makes the image of the canonical
half-disk into a wedge with the angle at $w_n=0$ equal to $2\pi/n$.
Figure (\ref{f3}-c) shows the disk $D_0$ in the $w_n$-plane with
the puncture and the local
coordinate. For any fixed $n$ we call the $\langle n|$ the
resulting surface state. Thus we have
\be \label{edefn}
\langle n| \phi\rangle \equiv \langle \ff_n\circ \phi(0)\rangle_{D_0}\,
\qquad \forall \phi
.
\ee

\medskip
The state obtained when $n=1$ is the
identity state (see Figure \ref{f4}-a). For this state the local
coordinate patch  
in the $w_n$ plane 
covers 
the full unit disk $D_0$ with a
cut on the
negative real axis. The left-half and the
right-half of the string coincide 
along this cut. The state $n=2$ is the vacuum state. In this case the
image of $H_U$ covers the right half of the full unit
disk $D_0$ in the $w_n$ plane. 
In the $n\to \infty$ limit, the image of $H_U$ 
in the $w_n$ coordinate is  a `thin sliver' of the  disk $D_0$
(Figure \ref{f4}-b). 
It was seen in \cite{0006240} that the limit $n\to \infty$ of
$\langle n|$ gives
rise to a well-defined state. In the next subsection we shall give a
detailed explanation 
for this result.
This surface state $\bsliv$, called the sliver, has the
property that the left-half and the right-half of the
string are as far as they can be on the unit disk.

Next we  describe the state $|n\rangle$ 
using the coordinate $\gl$ on
the upper half plane (see Fig.\ref{f3}-d). We have 
\be
\label{ncoor}
\gl_n = h^{-1} (w_n)  = i \,\,{1-w_n\over 1+ w_n} = \tan \Bigl ( -
{i\over 2}
\ln w_n \Bigr)\, .
\ee
The composition of \refb{ef1} and \refb{ncoor} gives
\cite{0006240}
\be
\label{e2}
\gl_n = \tan \Bigl (\, {2\over n} 
\tan^{-1} (\lo)\Bigr) \equiv \wt f_n (\lo) \, ,
\ee
and we have
\be \label{ed6}
\langle n|\phi\rangle = \langle \wt f_n\circ 
\phi(0)\rangle_{D_H}\, .
\ee

Finally we introduce the coordinate $\gf$ for the
presentation of the surface state
$|n\rangle$ using a disk $\wh D$. 
 Using eqs.\refb{ed4} and
\refb{ih} we see that
\be \label{ed7}
\gf_n = (w_n)^{n/2}\, .   
\ee
This is simply the map inverse to that taking
Fig.\ref{f3}-b to Fig.\ref{f3}-c, but extended for
all of $D_0$.  Under this map
the unit disk
$D_0$ in the
$w_n$-coordinates is mapped to a cone in the $\gf_n$ coordinate, 
subtending an
angle $n\pi$ at the origin $\gf_n=0$. We
shall denote this cone by $\wh D_n$.
 Thus we have
\be \label{ed8}
\langle n|\phi\rangle = 
\langle h\circ \phi(0)\rangle_{\wh D_n}\, .
\ee
In the $n\to\infty$ limit $\wh D_n$ can be viewed as an infinite helix,
as 
shown in Fig.\ref{f9}.

\subsection{Using SL(2,R)-to resolve singular coordinates}

All three descriptions of the sliver, using eqs.\refb{edefn}, \refb{ed6}
and
\refb{ed8}, are apparently singular. \refb{edefn} and \refb{ed6} are
singular since the corresponding maps $\ff_n(\lo)$ 
and $\wt f_n(\lo)$,
defined in eqs.\refb{ih} and \refb{e2} are singular in the $n\to\infty$
limit. On the other hand, \refb{ed8} is singular since the region $\wh
D_n$ becomes a cone with infinite excess angle at the origin in the
$n\to\infty$ limit. We shall now show that these are only apparent
singularities, and that for any Fock space state $|\phi\rangle$,
$\bsliv\phi\rangle$ is a perfectly finite number.

\begin{figure}[!ht]
\leavevmode
\begin{center}
\epsfxsize = 15 cm \epsfbox{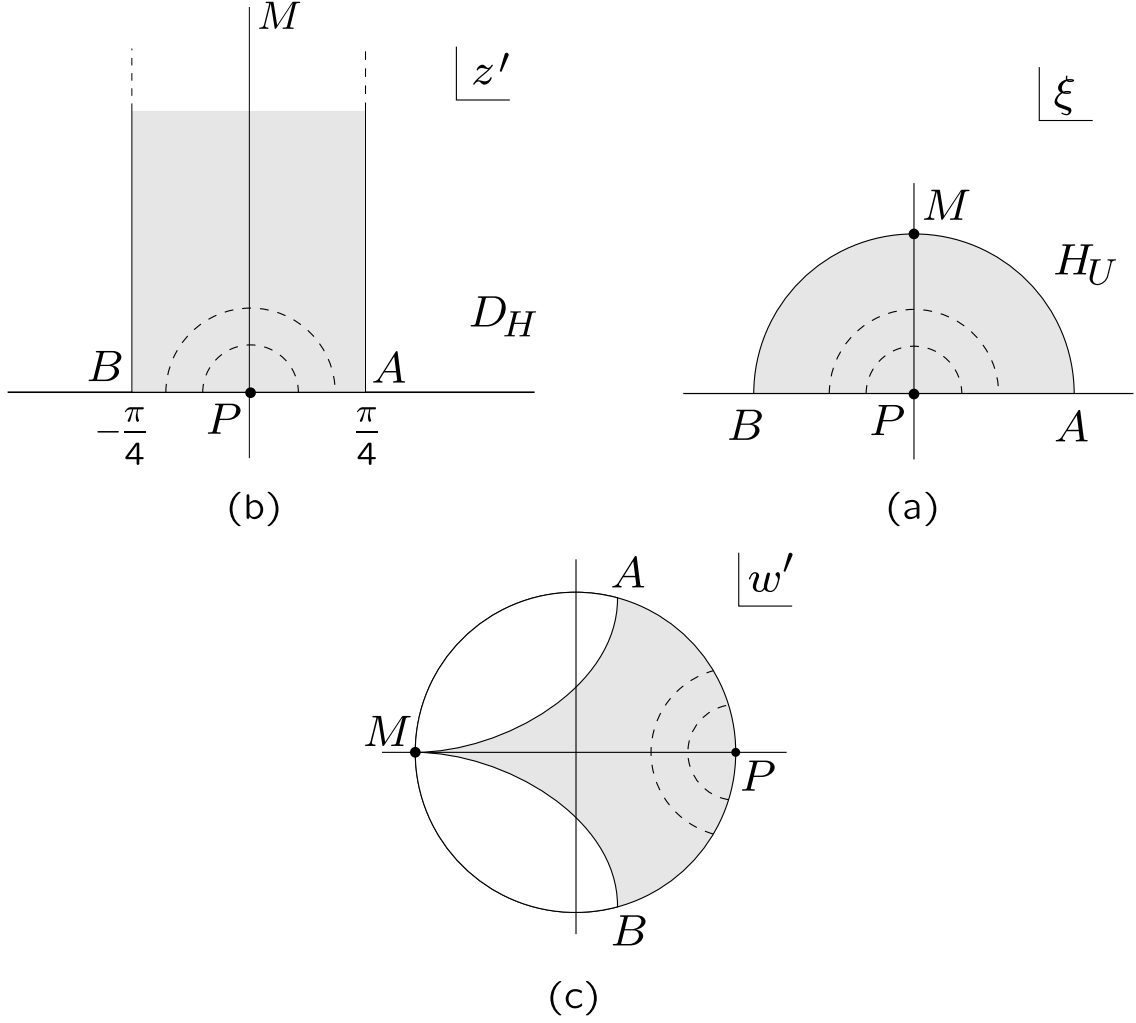 }
\end{center}
\caption[]{\small  The regular presentations of the sliver
surface state  with  non-singular coordinates at the
puncture.  In (b) the disk $D$ 
is the full upper half plane,
and the image of the local coordinate disk $H_U$ is the vertical strip
whose boundaries are the left-half and  right-half of the string. The
string midpoint is at  infinity. (c)  
Mapping $D_H$ 
back to a unit
disk with $w' = h(z')$
we find a regular presentation on the disk.} \label{f5}
\end{figure}

The main ingredient of the proof will be the SL(2,R) invariance of the
correlation functions in the upper half plane. Given any SL(2,R) map
$R(z)$, we have the relation:
\be \label{esl2r}
\langle \prod_i \OO_i(x_i)\rangle_{D_H} = \langle \prod_i
R\circ \OO_i(x_i)\rangle_{D_H}\, ,
\ee
for any set of operators $\OO_i$ and with $D_H$ denoting the upper half
plane.
Thus we can rewrite equation
\refb{ed6} as:
\be \label{ed6a}
\langle n|\phi\rangle = \langle R_n\circ \wt f_n\circ
\phi(0)\rangle_{D_H}\, ,
\ee
for any set of SL(2,R) maps $R_n$. Thus if we can find a sequence of maps
$R_n$
such that $R_n \circ \wt f_n$ approaches a finite limit $f$ as
$n\to\infty$ with $f(\lo)$ non-singular at the origin, then we can
define the sliver $\bsliv$ through the relations:
\be \label{eslivnewdef}
\bsliv\phi\rangle = \langle f\circ
\phi(0)\rangle_{D_H}\, .
\ee
In this case we choose:
\be \label{enn1}
R_n(z)={n\over 2} z\, ,
\ee
so that 
\ben \label{enn2}
f(\lo) &=& \lim_{n\to\infty} R_n\circ \wt f_n(\lo)\nonumber\\
& =&
\lim_{n\to\infty} {n\over 2} \tan\Big({2\over  n}\tan^{-1}(\lo)\Big) =
\tan^{-1} \lo\, .
\een
Since this map is non-singular at $\lo=0$, this provides a finite answer
for
$\bsliv\phi\rangle$ for any Fock space state $|\phi\rangle$, thereby
providing a non-singular description of the sliver. We shall denote by
\be \label{disp}  
\gl'=f(\lo)\, ,
\ee
the new coordinate on $D_H$.
This picture of the sliver state is shown in Figure \ref{f5}(a,b). The
local coordinate patch  
has become the full strip bounded by
the lines $\Re (\gl') = \pm \pi/4$. These two semiinfinite lines are
the left- and right- halves of the string. They meet at the midpoint $M$
which has been sent to $\gl'= i\infty$. This is actually the only
reminder that the sliver state is a limit of a sequence of fully
regular states. While the coordinate at the puncture can be taken
to be regular at this limit, and thus the state $\langle \Xi|$
is well-defined, in this limit the map of $H_U$ into the disk
fails to be regular at one point. The string midpoint has been
sent to the boundary of the disk.
While this does not affect any computation involving
Fock space states, this fact is
significant in that it shows that {\it there is no fully regular
geometric presentation of the sliver}. 
This might cause some worry since, as mentioned before,
for gluing purposes the  image of the arc
$\{ |\xi| = 1,~ 0<\arg (\xi) < \pi\}$  should be inside the
disk. We shall see, however, that gluing operations involving the sliver
can be made well-defined despite this singularity.

\begin{figure}[!ht]
\leavevmode
\begin{center}
\epsfxsize = 15 cm \epsfbox{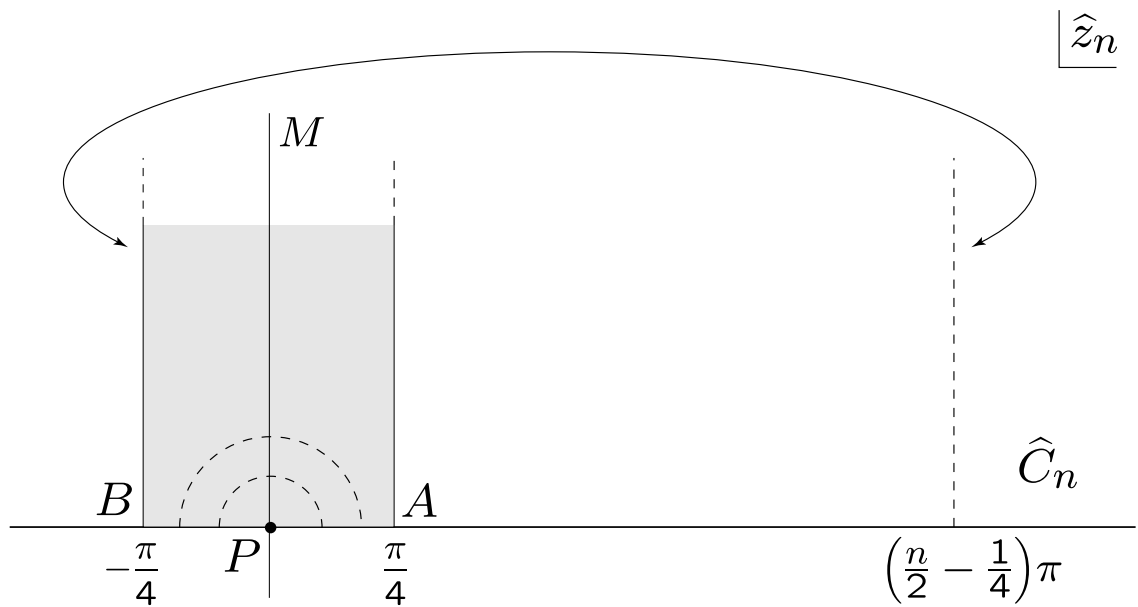 }
\end{center}
\caption[]{\small The finite $n$ approximation
to the sliver presented in the $\wh z_n$ coordinate. 
} \label{f10}
\end{figure}

This description of the sliver can be obtained by considering another
sequence of maps. For this we introduce a new coordinate system
\be \label{ecor1}
\wh z_n = {1\over 2 i} \ln \gf_n\, .
\ee
The cone $\wh D_n$ in the $\gf_n$ coordinate system maps to a
semiinfinite cylinder
$\wh C_n$ in the $\wh z_n$ coordinate system with $\wh z_n$ spanning the
range:
\be \label{ecor2}
-{\frac{\pi}{4}} \le \Re(\wh z_n) \le (\frac{n}{2}-{1\over 4}) 
\pi\, , \quad
\Im(\wh z_n)\ge 0, \quad \wh z_n \simeq \wh z_n + n\,{\pi\over  2}\, .
\ee
The local coordinate patch 
is the region:
\be \label{ecor3}
-{\pi/4} \le \Re(\wh z_n) \le{\pi/4} , \quad \Im(\wh z_n)\ge 0\, .
\ee
This has been shown in Fig.~\ref{f10}.
The relationship between $\wh z_n$ and the local coordinate $\xi$ follows
from eqs.\refb{ed4} and \refb{ecor1}:
\be \label{ecor5}
\wh z_n =\tan^{-1}\xi  \equiv  f(\xi)\, .
\ee
Thus we have
\be \label{ecor6}
\langle n| \phi\rangle = \langle 
f\circ \phi(0)\rangle_{\wh C_n} \quad
\forall
|\phi\rangle\in\HH\, .
\ee
Note now that using the periodicity along the $\Re(\wh z_n)$ direction
we
could take the range of $\Re(\wh z_n)$ to be $-n\pi/4\le \Re(\wh z_n)\le
n\pi/4$. In this case as $n\to\infty$, $\wh C_n$ approaches 
the full UHP and the
coordinate $\wh z_n$ approaches
the coordinate $z'$ introduced earlier.

Having obtained a regular description of the sliver we can map
it back to the unit disk by writing
$w' = h (\gl')$, with $h$ defined in \refb{ed1}. This gives us
Figure \ref{f5}-c; the regular presentation in the disk. The only
remnant of the singular origin is the fact that the open string
midpoint  reaches the boundary of the disk
and the open string 
develops a cusp at the midpoint. 
As a result of this, at least intuitively, 
the surface representing the sliver appears to be cut into two 
disjoint pieces by the local coordinate patch. One could therefore
expect the wavefunctional of the sliver to factorize into two
wavefunctionals, one corresponding to the data on the left-half
of the string, and the other to the data on the right-half of
the string.  It was seen  in refs.~\cite{TWO,gross-taylor}
that this intuitive   expectation is indeed realized.  

\begin{figure}[!ht]
\leavevmode
\begin{center}
\epsfxsize = 15 cm \epsfbox{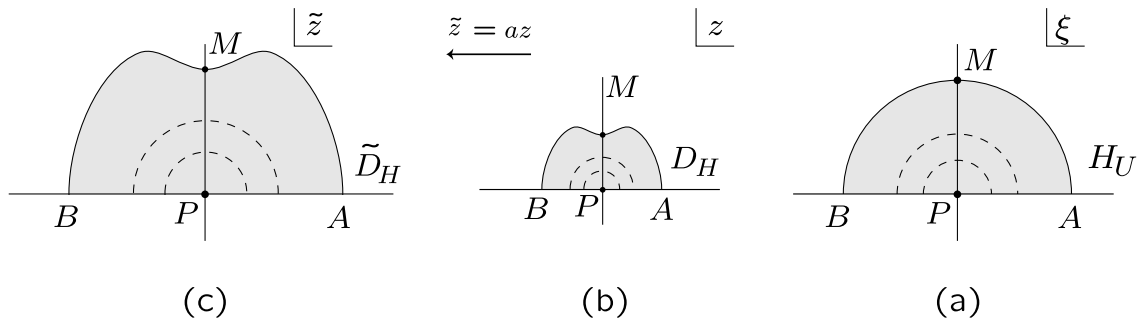 }
\end{center}
\caption[]{\small The power of conformal isometries in
resolving singular coordinates on once-punctured disks
is illustrated by showing two versions of a disk ($D_H$
and $\tilde D_H)$ that, while
looking different, 
represent the same Riemann surface.
Using an  $a\to \infty$ limit one could resolve a vanishingly
small image of $H_U$ in $D_H$.
} \label{f6}
\end{figure}

\medskip
We would like to emphasize that the use of SL(2,R) invariance
to resolve singular coordinates is 
most effective 
for once punctured disks. For the
case of three punctured disks, the three (distinguishable) punctures can
always be fixed at three points and there are no
SL(2,R) invariances left. If the local coordinate at one puncture
is singular, nothing can be done about it.   This is not the case for
once punctured disks, because after
fixing the position of the puncture there are powerful conformal
isometries left over.\footnote{  
This of course does not mean that any singularity of the 
local coordinate patch can be removed by such transformations, but there
is a reasonable degree of freedom.
} For example, after fixing the
puncture at
$z =0$, the maps
$\tilde z = a z/(c z + d)$ 
do change the looks of the local
coordinate as shown in Figure \ref{f6}. Even though the 
local coordinate patch in the 
$z$ or $\tilde z$ upper half planes look different,
and even though the functions $\tilde z = \tilde g(\xi)$ and
$z = g (\xi)$ are different, as Riemann surfaces with
local coordinates they are indeed {\it identical}.
For the case of two punctured disks, with punctures fixed
at $z=0$ and $z=1$, for example, the maps $\tilde z = az/(z + (a-1))$
preserve the punctures. Near $z=0$ the map looks like $\tilde z =
az/(a-1)$ and near $z=1$ it looks like $\tilde z = 1 + (z-1) (a-1)/a$.
Since the scaling factors are inverses of each other, if only one
coordinate is singular it cannot be resolved. Only particular
cases when the coordinates are singular in a related way could be
resolved.

\subsection{Star multiplication of surface states}  

\begin{figure}[!ht]
\leavevmode
\begin{center}
\epsfxsize = 15 cm \epsfbox{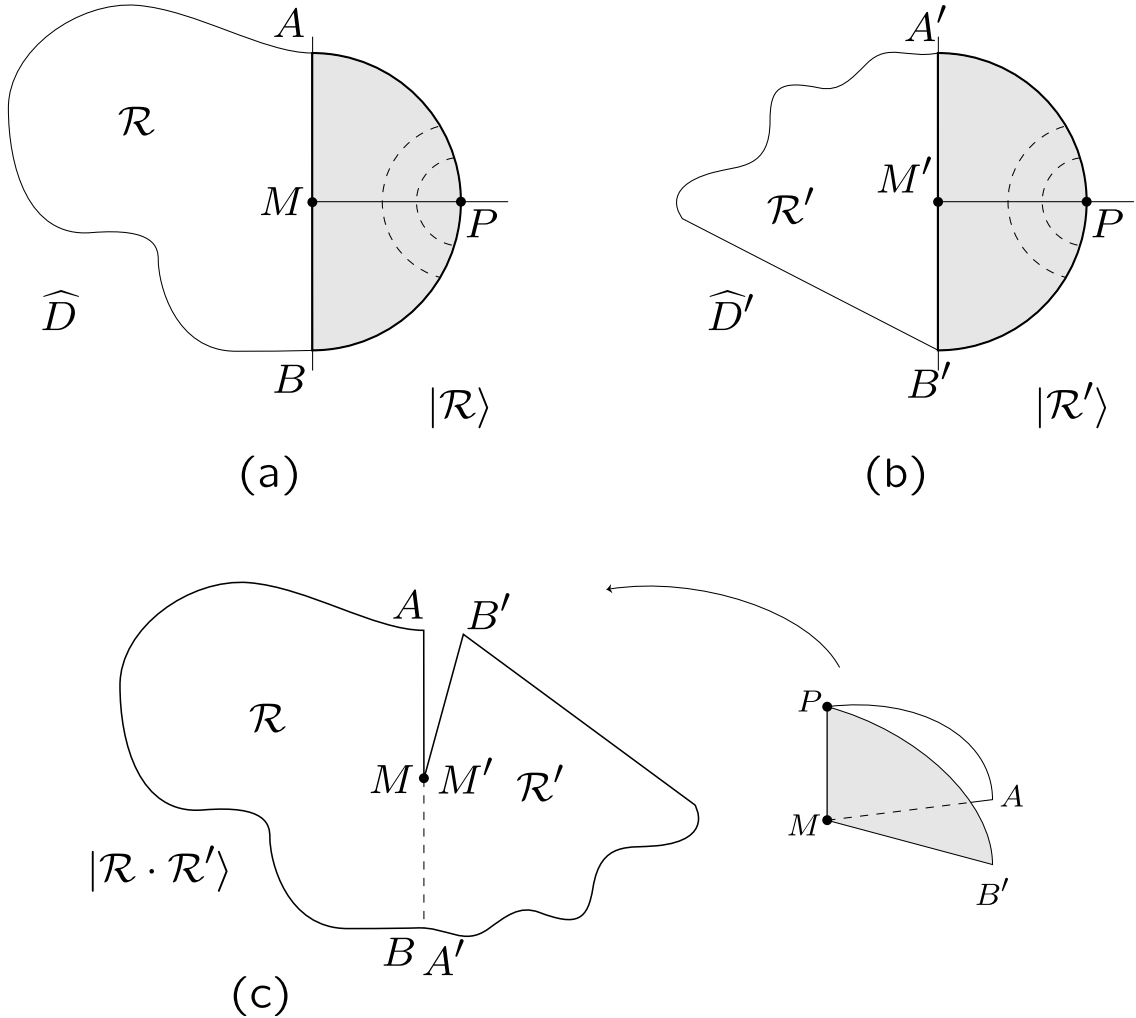 }
\end{center}
\caption[]{\small  Star multiplication of surface
states is simple if the states are presented with $\wh w$
coordinates, in which case the local coordinate appears
as a (shaded) half-disk in the disk $\wh D$.  Here are shown
two surface states; $|\RR\rangle$ in  (a), and $|\RR'\rangle$
in (b). The star product is shown in (c) and is obtained gluing
the regions $\RR$ and $\RR'$ as indicated, and attaching 
the half disk (shown to the right) representing the local
coordinate and the puncture. 
} \label{f7}
\end{figure}

The star multiplication of two surface states is easiest to describe by
representing them in 
the $\gf$ coordinate system. Let us assume that we have two such surface
states, and associated with them are the regions $\wh D$ and $\wh D'$,
which describe the images of the disk in the corresponding $\gf$
coordinate system. Each of these disks contain as a subspace the region
$h(H_U)$, $-$ the local coordinate patch. 
We shall denote by $\RR$ and $\RR'$ the
complement of
$h(H_U)$ in
$\wh D$ and $\wh D'$ respectively. 
By an abuse of notation we shall denote
the surface states associated with the disks $\wh D$ and $\wh D'$ by 
$|\RR\rangle$
and $|\RR'\rangle $ respectively.
This is represented diagrammatically in Figure \ref{f7},
where in parts (a) and (b) we show two surface states $|\RR\rangle$
and $|\RR'\rangle $ built in this way. The regions $h(H_U)$ have been
shown shaded in this figure.
In fact, once we have made it clear that the local coordinate is
presented in the specific fashion chosen here, we could simply represent
pictorially the surfaces as only the regions $\RR$, namely
the image of the
full disk minus the local coordinate  
patch.\footnote{Leaving off
the local coordinate, of course, carries a small risk of confusion,
especially because the local coordinate patch 
could be
presented in
other canonical ways, for example, as semi-infinite strips.}

If the surface states are presented this way, the star multiplication
$|\RR\rangle * |\RR'\rangle$ is readily performed. One removes the
local coordinate patches 
from the disks $\wh D$ and $\wh D'$, and glues
the right half (as viewed from the region of 
$\wh D$ {\it outside} the local coordinate patch) 
$MB$ of the
$\RR$ open string to the left half $A'M'$ of the
$\RR'$ open string. The result is the composite
region $\RR\cdot \RR'$ shown in the figure. This region represents
the star product of the states, namely
\be
|\RR\rangle * |\RR'\rangle = | \RR \cdot \RR'\rangle\,.
\ee
In this region, the string is $AMB'$, with $AM$ the left half, and
$MB'$ the right half. The local coordinate patch 
shown to the right,
must
be glued in to produce the full picture of the surface state.

\begin{figure}[!ht]
\leavevmode
\begin{center}
\epsfxsize = 15 cm \epsfbox{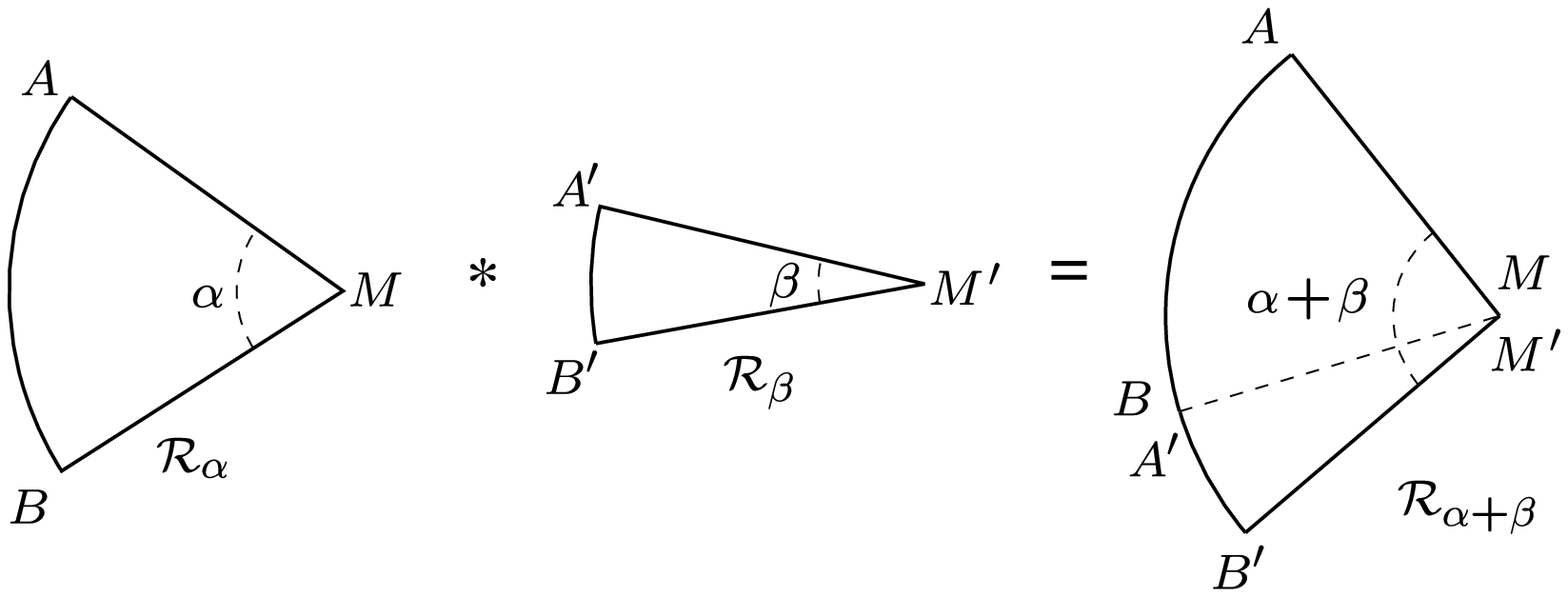 }
\end{center}
\caption[]{\small The star multiplication of a sector 
state with angle $\alpha$ to a sector state with angle
$\beta$ gives a sector state with angle $\alpha+ \beta$.
Sector states are just another presentation of wedge
states.
} \label{f8}
\end{figure}

We now discuss the multiplication of wedge states $ |n\rangle$.
As we have already discussed, in the $\wh w$ coordinate system relevant
for the state $\langle n|$, the disk becomes a cone subtending an angle
$\pi n$ at the origin. If we remove the local coordinate patch 
the left over region becomes a sector of
angle $\pi (n-1)$.  If we denote
by $| \RR_\alpha\rangle$ a {\it sector state} arising from a sector of
angle $\alpha$,  we have the identification of sector states with
wedge states: 
\be
\label{wis}
 |n\rangle = |\RR_{\pi(n-1)} \rangle \,.
\ee  
Moreover, it is clear from our discussion above that star multiplication
of sector states simply gives a sector whose angle is the sum of those
in the product:
\be
\label{hik}
|\RR_\alpha\rangle * |\RR_\beta\rangle = |\RR_{\alpha+ \beta}\rangle \,,
\ee
This is illustrated in Figure \ref{f8}.  Indeed the multiplication 
rule \refb{hik} is consistent with $|n\rangle * |m\rangle =
|n+m-1\rangle$, which is the familiar multiplication rule for
wedge states.\footnote{We hope the reader will  
 not be confused by
the dual use of the term surface state. In the {\it wedge states}, the
focus is on the local coordinate which looks like a wedge  inside a unit
disk, in the  {\it sector states}, the focus is on the complement of the
local coordinates in the disk when the local coordinate patch 
is presented as a half-disk.
This complement looks like a sector.} The sliver arises from
$n\to
\infty$ and therefore  corresponds to a sector state with infinite
angle! This state was shown schematically in Figure \ref{f9}.
Indeed, it should be noted that among sector states only two can multiply
to
themselves, according to \refb{hik}.  One is the zero-angle sector state
$|\RR_0\rangle$ which is the identity $|\II\rangle$ (or the wedge state
$|n=1\rangle$). The other is the sliver, which can be viewed as the zero
angle  wedge state,
or as the infinite angle sector 
state. Indeed, from \refb{hik} we have 
$|\RR_\infty\rangle * |\RR_\infty\rangle = |\RR_{\infty}\rangle$.

\medskip
For later use it will be useful to work out the precise relationship
between the different coordinate systems appearing in the description of 
the product state
$|\RR_{\alpha+\beta}\rangle$ and the states $|\RR_{\alpha}\rangle$ and
$|\RR_{\beta}\rangle$. Again this is simple 
in the $\gf$ coordinate
system.
For this let us take
\be \label{ealphamn}
\alpha = \pi(m-1), \qquad \beta=\pi(n-1), \qquad \alpha + \beta =
\pi(m+n-2)\, .
\ee
If we denote by $\gf_m$, $\gf_n$ and $\gf_{m+n-1}$ the
$\gf$ coordinates associated with the sector states
$|\RR_{\alpha}\rangle$,
$|\RR_{\beta}\rangle$ and $|\RR_{\alpha+\beta}\rangle$ respectively, we
have
\be \label{ecorpat}  
\gf_{m+n-1} = \cases{ \gf_m \quad \,\,\,\,\,\,
\hbox{in}\quad \RR_\alpha  \cr
 e^{i\alpha} \gf_n \quad \hbox{in}\quad \RR_\beta\,\,. }
\ee
In the $\wh z_n$ coordinate system introduced in eq.\refb{ecor1}
the gluing
relations \refb{ecorpat} take a very  simple form: 
\be \label{ecor4}  
\wh z_{m+n-1} = \cases{ \wh z_m 
\qquad\qquad\qquad \,\,\,\hbox{for} \quad {\pi\over 4}
\le
\Re(\wh z_m) \le (\frac{m}{2}-{1\over 4})\pi, \cr\cr
\wh z_n + {1\over 2} (m-1)\pi \quad \hbox{for} \quad
{\pi\over 4} \le
\Re(\wh z_n) \le (\frac{n}{2}-{1\over 4})\pi\, .}
\ee

\sectiono{General D-brane solutions}   
\label{s3}

In this section we  shall give 
deformations of the sliver to generate 
new solutions of the
equations of motion 
representing D-branes described by general BCFT's.
The general idea is simple.
We denote by BCFT$_0$ the {\it reference} BCFT  
in whose Hilbert space the
string field takes value.
The sliver of BCFT$_0$, 
whose matter part
represents the D-brane solution associated with BCFT$_0$, 
is  the surface  state described in the previous section,
with the  specific boundary condition corresponding to BCFT$_0$ on
the boundary of the surface.
To get a solution representing the single D-brane 
of  some other boundary conformal field theory 
BCFT$'$ we must represent the sliver of BCFT$'$ on the
state space of BCFT$_0$, as we now explain. 
The construction assumes that BCFT$_0$ and BCFT$'$ have the same bulk
conformal field theory,  but of course, differ  in
their boundary interactions.

 Usually the sliver $\Xi'$ of BCFT$'$ will be
described in the same manner as discussed in the last section, with all
the correlation functions now being calculated in BCFT$'$. This, however,
would express $\Xi'$ as a state in the Hilbert space $\HH'$ of BCFT$'$, 
since in eq.\refb{eslivnewdef}, for example, $\phi$ will now represent a
vertex  
operator of BCFT$'$. In order to express $\Xi'$ in the Hilbert space of
BCFT$_0$, we adopt the following procedure. 
As discussed in section \ref{sq1}, at 
an intuitive
level, the wave-functional for $\langle\Xi'|$ is a 
functional of $\vp_0(\sigma)$
with $\sigma={1\over
i}\ln\xi$ labeling the coordinate along the string,
obtained by 
functional
integration over the two dimensional fields $\vp$ on the full disk (UHP
in this
case) minus the local coordinate patch, with boundary condition
$\vp=\vp_0(\sigma)$ on the boundaries
$\Re(z')=\pm\pi/4$, and the boundary condition
appropriate to BCFT$'$ on the rest of the boundary of this region ($z'$
real,
$|\Re(z')|\ge \pi/4$). 
On the other hand given a state $|\phi\rangle$ in  
the Hilbert space $\HH$ of BCFT$_0$, 
we represent
the
wave-functional of
$|\phi\rangle$ as a functional of $\vp_0(\sigma)$,  obtained by
performing path
integral over $\vp$ in the inside of the  local coordinate patch {\it
with
boundary
condition appropriate for BCFT$_0$} on the real axis, 
the 
vertex operator 
$\phi$ of BCFT$_0$ 
inserted at the origin in the local coordinate system $\xi$, and 
the boundary condition $\vp=\vp_0(\sigma)$ on the semi-circle.
Finally in order to calculate $\langle\Xi'|\phi\rangle$ we
take
the product of the wave-functional of $\langle\Xi'|$
and the wave-functional of $|\phi\rangle$ and integrate over
$\vp_0(\sigma)$. The result will be a functional integral
over the fields $\vp$ on the full UHP, with boundary condition
appropriate to BCFT$'$ in the range $|\Re(z')|\ge \pi/4$,
boundary condition corresponding to BCFT$_0$    
for $|\Re(z')|\le \pi/4$ and the 
vertex operator
$\phi$ inserted at the origin {\it in the local coordinate system}.
This can be expressed as
\be \label{eexpress}
\langle\Xi'|\phi\rangle = \langle f\circ\phi(0)\rangle''\, ,
\ee
where $\langle ~\rangle''$ denotes  
correlation function in a theory where
we have BCFT$'$ boundary condition for $|\Re(z')|\ge \pi/4$ and BCFT$_0$
boundary condition for $|\Re(z')|\le \pi/4$. $\phi$ is a vertex operator
in BCFT$_0$ and $f(\xi)=\tan^{-1}\xi$ as usual.

In what follows, we shall show that after appropriate ultraviolet 
regularization,  
$\Xi'$ defined this way squares to   
itself under $*$-multiplication, and also has the right tension for
describing a D-brane associated to BCFT$'$. 
We will begin by considering this general case in 
the next subsection. 
Then we will discuss the situation where 
BCFT$'$ is replaced by a general two
dimensional field theory obtained from BCFT$_0$ by some boundary 
perturbation.

\subsection{Solution describing an arbitrary D-brane} \label{sn1}

We shall now describe the construction of the solution of the SFT 
equations of motion describing a D-brane 
corresponding to an 
arbitrary boundary conformal field theory BCFT$'$ with 
the same bulk  
CFT as BCFT$_0$.
We start with the definition of $\Xi'$ given in eq.\refb{eexpress}.
The effect of the change of the boundary condition beyond $|x'|\ge \pi/4$
can be represented by inserting boundary condition
changing vertex operator $\sigma^\pm$ (discussed {\it e.g.} in
\cite{cardy, 9704006}) 
at $x'=\pm\pi/4$. In other words we can express 
\refb{eexpress} as
\be \label{exy1}
\langle\Xi'|\phi\rangle = \Bigl\langle f\circ\phi(0) 
\, \sigma^+({\pi\over 4}) \sigma^-(-{\pi\over 4}) \Bigr\rangle\, .
\ee
If we denote by D and D$'$ the  D-branes
associated with BCFT$_0$ and BCFT$'$ respectively, then $\sigma^+$
denotes 
the vertex operator for the ground state of an open string whose left
end  (viewed from inside the UHP) 
is  on the D$'$-brane 
and right end is on the D-brane, whereas $\sigma^-$ denotes the vertex 
operator  for the ground state of open string whose left end is     
on the D-brane and right end is on the D$'$-brane. 
In anticipation of  short-distance divergences, we shall actually put
$\sigma^-$ and $\sigma^+$ at 
$(-{\pi\over 4}-\epsilon)$ and $({\pi\over 4}+\epsilon)$ respectively, 
where $\epsilon$ is a small positive number. We 
shall also use the description of the sliver as limit of finite $n$
wedge 
states in the $\wh z_n$ coordinate 
introduced in eq.\refb{ecor1}. Thus we have
\be \label{exy2}
\langle\Xi'|\phi\rangle =   \lim_{n\to\infty} 
\Bigl\langle f\circ\phi(0)
\, \sigma^+\Bigl({\pi\over 
4}+\epsilon\Bigr) \sigma^-\Bigl({n\over 2}\pi-{\pi\over 4}-\epsilon\Bigr)
\Bigr\rangle_{\wh  C_n}\, .
\ee   
We now 
calculate $\Xi'*\Xi'$.
{}From the gluing relations \refb{ecor4} we get,
\ben \label{exy3}  
\langle \Xi' * \Xi'|\phi\rangle &=& \lim_{m,n\to\infty}  \Biggl\langle 
f\circ
\phi(0) \,\, \sigma^+\Bigl( {\pi\over 4} +\epsilon\Bigr)
\,\sigma^-\Bigl( ({m\over 2}-{1\over 4})\pi 
-\epsilon\Bigr) 
\nonumber \\
&&  \sigma^+\Bigl(
({m\over
2}-{1\over 4})\pi+\epsilon\Bigr) 
\sigma^-\Bigl(
{1\over 2} (m+n-1) \pi -{\pi\over 4} 
-\epsilon\Bigr)\Biggr\rangle_{\wh C_{m+n-1}} \, . 
\een
Thus the $\sigma^\pm$ at ${1\over 2}
(m-{1\over  2})\pi 
\pm\epsilon$
come close as $\epsilon\to 0$ and give rise to a divergent factor 
$(2\epsilon)^{-2h}$ where $h$ is the conformal weight of $\sigma^\pm$.
Hence we have $\Xi' * 
\Xi'= (2\epsilon)^{-2h}\Xi'$. 
This requires us to
redefine $\Xi'$ by absorbing a factor of $(2\epsilon)^{2h}$,
so that it squares to itself under $*$-multiplication
\be 
\label{rt}
\Xi'_\rr 
\equiv (2\epsilon)^{2h}\,\, \Xi'  \quad \to \quad
\Xi'_\rr * \Xi'_\rr = \Xi'_\rr \,.
\ee
We note, however, that even for finite $\epsilon$, the state
$\Xi'_\rr$ still squares to itself. 
Indeed, the
product
$\sigma^+({1\over  2}(m-{1\over 2})\pi + \epsilon) \sigma^-( 
{1\over
2}(m-{1\over 2})\pi -\epsilon)$ 
can be expanded in an operator product 
expansion,
and since these operators are moved to $\infty$ in the $m,n\to\infty$
limit, only the identity operator in this operator product expansion 
would contribute. 
The coefficient of the identity operator is given by $(2\epsilon)^{-2h}$ 
even for finite $\epsilon$.

Since BCFT$_0$ and BCFT$'$ differ only in the matter sector, it is clear 
that 
$|\Xi'_\rr\rangle$ has the factorized form
\be \label{epfac}
|\Xi'_\rr\rangle=|\Xi_g\rangle \otimes |\Xi'_{\rr,m}
\rangle\, ,
\ee
where $\Xi_g$ is a universal ghost factor. Normalizing
$\Xi_g$ 
(which is independent of 
the choice of BCFT$'$) 
such that $\Xi_g*^g\Xi_g=\Xi_g$,
we can
ensure that
\be \label{eprodm} 
\Xi'_{\rr,m} *^m \Xi'_{\rr, m} = \Xi'_{\rr,m}\, .
\ee
Thus we can now construct a new D-brane solution by taking the product
$|\Psi_g\rangle\otimes |\Xi'_{\rr,m}\rangle$, where 
$|\Psi_g\rangle$
is the same universal ghost state that appears in the construction of the
D-brane solution corresponding to BCFT$_0$.

We shall now calculate the tension associated with this new D-brane 
solution. For this we need to compute
$\langle
\Xi_{\rr,m}'|\Xi_{\rr,m}'\rangle_m$, where the 
subscript $m$ denotes matter. 
We have,
\be \label{exy4}
\langle\Xi_{\rr,m}'|\Xi_{\rr,m}'\rangle_m = (2\epsilon)^{4h}\,\,
\langle\Xi_{m}'|\Xi_{m}'\rangle_m \,.
\ee
Calculation of $\langle\Xi_{m}'|\Xi_{m}'\rangle_m$  is again simple
in the $\wh z_n$
coordinate system. We first compute the $*$-product of the two states,
and then in the final glued surface with coordinate $\wh z_{m+n-1}$ we
remove the 
local coordinate patch 
$-\pi/4\le \Re(\wh
z_{m+n-1})\le \pi/4$ and identify the lines $\Re(\wh z_{m+n-1})=\pm
\pi/4$. This produces the semi-infinite cylinder $\wt C_{m+n-2}$
defined by  
${\pi\over 4} \leq \Re(
z_{m+n-2}) \leq {\pi\over 4}+ (m+n-2){\pi\over 2}$ and $\Im (z_{m+n-2})
\geq 0$. 
We therefore find 
\ben \label{exy5}
\langle\Xi_{m}'|\Xi_{m}'\rangle_m 
&=& \lim_{m,n\to\infty}  \Biggl\langle 
\sigma^+\Bigl( {\pi\over 4} +\epsilon\Bigr)\,\,
\sigma^-\Bigl( ({m\over 2}-{1\over 4})\pi 
-\epsilon\Bigr) 
\nonumber \\
&&  \sigma^+ \Bigl( {m\over
2}-{1\over 4})\pi + \epsilon\Bigr)  
\sigma^-\Bigl(
{1\over 2} (m+n-1) \pi -{\pi\over 4} 
-\epsilon\Bigr)\Bigg\rangle_{\wt C_{m+n-2}} \, , 
\een
where the correlation function is now being computed in the matter BCFT.
We get 
a factor of
$(2\epsilon)^{-2h}$ 
coming from $\sigma^\pm$ inserted at $({m\over 2}-{1\over 4})\pi  
\pm\epsilon$ as before, but there is 
another factor of $(2\epsilon)^{-2h}$
coming from the other $\sigma^\pm$ insertions that happen at points
separated by a (minimal) distance $2\epsilon$ on the cylinder
$\wt C_{m+n-2}$.
These exactly cancel the explicit
factor of
$(2\epsilon)^{4h}$ in \refb{exy4}.
Since from the definition 
of $\sigma^\pm$ it is clear that in the $\epsilon\to 
0$ limit we have BCFT$'$ boundary condition on the full real 
$z_{m+n-2}$ 
axis, we 
find that $\langle \Xi'_{\rr,m} |
\Xi_{\rr, m}'\rangle_m$ is given by the partition function of the
deformed boundary CFT on the $\wt C_{m+n-2}$  cylinder 
\be \label{exy6}
\langle
\Xi_{\rr,m}'|\Xi_{\rr,m}'\rangle_m = Z_{\wt C_{m+n-2}}
 (\hbox{BCFT}')
\sim Z_{D_0} (\hbox{BCFT}')\,,
\ee
where in the last step we relate this partition function to
the one on the standard unit disk. This is possible because of conformal
invariance. 
Any constant multiplicative factor that might appear due to conformal 
anomaly depends only on the bulk central charge and is independent of
the 
choice of BCFT$'$. This can at most give rise to a universal
multiplicative 
factor.
Since the partition function of BCFT$'$ on the unit disk is proportional
to  the tension of the corresponding 
D-brane~\cite{9511173,9707068, 9807161,9909072,0101200}, 
we see that the tension $\langle
\Xi_{\rr,m}'|\Xi_{\rr,m}'\rangle_m$ computed from  vacuum 
string field theory agrees with the known tension of the BCFT$'$
D-brane, up to an overall 
constant factor independent of BCFT$'$. 

\medskip
Arguments similar to the 
one given for $\Xi'*\Xi'$ show 
that the 
result \refb{exy6} holds even when $\epsilon$ is finite. In this case we 
have two pairs 
of $\sigma^\pm$ on the boundary, with the first pair being 
infinite 
distance away from the second pair. Thus we can expand each pair using 
operator product expansion 
and only the identity operator contributes, 
giving us back the partition function of BCFT$'$ on the disk.
{}From this we see that we have a one parameter family of solutions, 
labeled by $\epsilon$, describing the same D-brane. We expect these 
solutions to be related by gauge 
transformations, 
since $\p_\epsilon\Xi'_{\rr,m}$ has finite norm (as can be easily
verified) and hence is pure gauge according to the arguments to be 
given in section \ref{s6}.

\subsection{Multiple D-branes and coincident D-branes}
\label{snew4.2}

We first consider the construction of a configuration
containing various D-branes associated to different
BCFT$'$s.
To this end, we note that the star product $\Xi'*\Xi$ of
the BCFT$'$ solution and the BCFT$_0$ solution vanishes.
Indeed, using the same methods as in the previous
subsection, the computation of 
$\Xi'*\Xi$ leads to the cylinder $\wh C_{m+n-1}$ with a
$\sigma^+$ insertion at
${\pi\over 4}+\epsilon$ 
and a $\sigma^-$ insertion at $({m\over 2} -{1\over 4})-\epsilon$. In
the  $m,n\to\infty$ limit, $\sigma^-$ moves 
off to infinity and as a result
the  correlation function vanishes since $\sigma^-$ has 
dimension larger than zero {\em   as 
long as BCFT$_0$ and BCFT$'$ are different.} Similar arguments 
show that 
$\Xi*\Xi'$ and $\langle\Xi_m|\Xi'_m\rangle_m$ also vanish. Thus 
the matter part  of $\Xi+\Xi'_\rr$ is a new solution  
describing the superposition of the D-branes corresponding  to 
BCFT$_0$ and BCFT$'$. 
Since no special assumptions were made about BCFT$_0$ nor
BCFT$'$, it follows that 
 $\Xi'*\Xi''=\Xi''*\Xi' = 0$ and 
$\langle\Xi'_m|\Xi''_m\rangle_m =0$ for any two different 
BCFT$'$ and
BCFT$''$, and hence we can superpose any number of slivers to
form a solution. This in 
particular also includes theories which differ from each other by a
small 
marginal deformation. Special cases of 
this phenomenon, in the case of D-branes in flat space-time, have been 
discussed in ref.\cite{gross-taylor}.

\medskip
This procedure, however, is not suitable for superposing
D-branes 
associated with the same BCFT, {\it i.e.} for parallel coincident 
D-branes. For example,
 if we take BCFT$'$ to differ from BCFT$_0$ by
an  exactly 
marginal deformation with deformation parameter $\lambda$, 
then in the 
$\lambda\to 0$ limit the operators $\sigma^\pm$ 
both 
approach the identity operator (having vanishing conformal weight),
and although the argument
 for the  vanishing of $\Xi*\Xi'$ holds for any non-zero $\lambda$,
it breaks down at $\lambda=0$.

In order to construct a superposition of identical D-branes, one can
proceed in a different way. 
First consider getting coincident BCFT$_0$ branes. To this
end we introduce a modified BCFT$_0$ sliver
\be \label{exy222}
\langle\Xi_\chi|\phi\rangle =   \lim_{n\to\infty} 
\Bigl\langle f\circ\phi(0)
\, \chi^+\Bigl({\pi\over 
4}+\epsilon\Bigr) \chi^-\Bigl({n\over 2}\pi-{\pi\over 4}-\epsilon\Bigr)
\Bigr\rangle_{\wh  C_n}\, .
\ee   
Here  $\chi^\pm$ are a conjugate pair\footnote{We
need to choose $\chi^\pm$ to be conjugates of each other so that
the string field is hermitian.} of operators of BCFT$_0$,
having a common dimension $h$ greater than zero, and representing some 
excited states of the open string with
{\it both
ends having BCFT$_0$ boundary condition}. Thus, throughout the
real line we have BCFT$_0$ boundary conditions. We require
that the coefficient of the identity in the OPE $\chi^-(x) \chi^+(y)$ is
given by $ |x-y|^{-2h_i}$, and that this 
OPE does not
contain any  other operator of dimension $\le 0$.   

The clear parallel between eqn.~\refb{exy222} and eqn.~\refb{exy2},
describing the BCFT$'$ D-brane, implies that an  analysis
identical to
the one carried out in the previous section will show that:
\begin{enumerate}
\item This new state $\Xi_\chi$ (after suitable renormalization as in
eq.\refb{rt}) squares to itself under
$*$-multiplication.
\item The BPZ norm of the matter part of 
$\Xi_\chi$ is proportional to the
partition
function of BCFT$_0$ on the unit disk.
\item $\Xi_\chi$ has vanishing $*$-product with $\Xi$.
\end{enumerate}
Thus the matter part of this state gives another representation of the
D-brane associated with BCFT$_0$, 
and we can construct a pair of D-branes
associated with BCFT$_0$ by superposing the matter parts of $\Xi$ and
$\Xi_\chi$. 

This construction
can be easily generalized to describe 
multiple BCFT$_0$ D-branes. 
We construct different
representations of the
same D-brane by
using different vertex operators $\chi^{(i)\pm}$ in
BCFT$_0$ satisfying the `orthonormality condition' that
the coefficient of the identity operator in the OPE of
$\chi^{(i)-}(x) \chi^{(j)+}(y)$ is given by
$\delta_{ij} |x-y|^{-2h_i}$, and  that this 
OPE does not
contain any other operator of dimension $\le 0$. 
The correponding
solutions
$\Xi_{\chi^{(i)}}$ all have vanishing $*$-product with each other,
and hence can be superposed to represent multiple D-branes
associated
with BCFT$_0$.

If instead we want to construct a
superposition of identical BCFT$'$ D-branes, we need to
replace $\sigma^\pm$ in eq.\refb{exy2} 
by another pair of vertex operators
$\tau^\pm$ which represent some excited states of open strings with one
end satisfying BCFT$_0$ boundary condition and the
other end satisfying BCFT$'$ boundary condition. The corresponding state
$\Xi'_\tau$ will give
another representation of the D-brane associated with BCFT$'$, having
vanishing $*$-product with $\Xi'$ as long as  the
operator
product of $\sigma^-$ with $\tau^+$ does not contain the identity
operator or any other operator of dimension $\le 0$. Hence we can
superpose these solutions to construct new
solutions.

This procedure of adding vertex operators near $\pm\pi/4$ to create new
solutions representing the same D-brane is
the BCFT version 
of the use of excited states of half-strings \cite{TWO,gross-taylor} 
for the same purpose.

\subsection{Finite deformations of the sliver} \label{s4a}

We shall now consider a class of solutions associated with the 
sliver for boundary field theories which are not necessarily conformal.
Let us begin with the description of the sliver of BCFT$_0$ given in the 
$\wh z_n$
coordinate system defined in eq.\refb{ecor1}. We have from
eq.\refb{ecor6}:
\be \label{epp1}
\langle \Xi| \phi\rangle = \lim_{n\to\infty} \langle f\circ
\phi(0)\rangle_{\wh C_n} \quad
\forall
|\phi\rangle\in\HH\, ,
\ee
with $f(\xi)=\tan^{-1}\xi$, and $\langle\cdot\rangle$ denotes
correlation 
function calculated in BCFT$_0$. Now suppose $V$ is a local vertex 
operator in the matter 
sector  of BCFT$_0$. We define a new state $\langle \Xi^{V,\lambda}|$ 
through the
relation:\footnote{A construction 
that is similar in spirit but uses a
different geometry was suggested in ref.\cite{0011009}.}
\be \label{epp2}
\langle \Xi^{V,\lambda}| \phi\rangle = \lim_{n\to\infty} \Big\langle 
\exp\Big(-\lambda \int_{{\pi\over 4}}^{({n\over 2} -{1\over
4})\pi}
V(x_n)
dx_n\Big)
f\circ
\phi(0)\Big\rangle_{\wh C_n} \quad
\forall
|\phi\rangle\in\HH\, ,
\ee
where $x_n=\Re(\wh z_n)$, $\lambda$ is a constant, and the integration is
done over the real $\wh z_n$ axis {\it excluding the part that is inside
the local coordinate patch.}  
The description of this state can be made
more transparent by taking the full range of $\Re(\wh z_n)$ to be
$[-{n\over 4}\pi, {n\over 4}\pi]$ with the local coordinate patch
in the region $-{\pi\over 4}\le \Re(\wh z_n)\le{\pi\over 4}$, and
then
identifying $\wh z_n$ as the coordinate $z'$ on the upper half plane in
the $n\to\infty$ limit. This gives
\be \label{epp22}
\langle \Xi^{V,\lambda}| \phi\rangle = \Bigl\langle 
\exp(-\lambda \int_{|x'|>{\pi\over 4}}
V(x') \, dx') f\circ \phi(0)\Bigr\rangle_{D_H} \, ,  
\ee
where $x'=\Re(z')$. 
This expression should be treated  
as a correlation function in a theory where on part of the boundary we
have the usual boundary action corresponding to 
BCFT$_0$, 
and on part
of the boundary we have a modified boundary action obtained by adding
the integral of $V$ to the original action. In defining this we need
to use
suitable operator ordering, and regularization and renormalization
prescriptions 
to remove the
short distance singularities (see {\it e.g.} 
ref.\cite{9811237} for a discussion of this). 
Typically this requires $V$
to be of
dimension $\le 1$.  
In special cases, as discussed in
\cite{9303067},  
we may
also be able to include operators of higher dimensions, but as we shall
see later these do not generate new solutions. 
We shall not explicitly take into account the effects of ultra-violet 
divergences in our analysis, and hence the results of this and the next 
subsection will be somewhat formal. The analysis, however,
can be made
concrete in  specific examples, {\it e.g.} of the kind discussed in 
refs.~\cite{9303067}.  
If we regard the effect of $\lambda$ deformation to be
a change in the boundary condition, then \refb{epp22} represents a
correlation function with BCFT$_0$ 
boundary condition in the range
$-{\pi\over 4}\le x\le {\pi\over 4}$ and the modified boundary condition
outside this range.

We shall now show that $|\Xi^{V,\lambda}\rangle$ satisfies the relation:
\be \label{epp3}
\Xi^{V,\lambda} * \Xi^{V,\lambda} = \Xi^{V,\lambda}\, .
\ee
To compute the $*$-product in the left hand side of the above equation,
we use the $\wh z_n$ coordinate system, keeping $n$ finite and taking
the
$n\to\infty$ limit at the end. The advantage
of using the $\wh z_n$
coordinate system is in the simplicity of the gluing relations
\refb{ecor4};
since the coordinate $\wh z_{m+n-1}$ of the surface after gluing is
related to the coordinates $\wh z_m$ and $\wh z_n$ of the original
surfaces by simple shift, there is no conformal transformation of the
factors $\int V(x_n) dx_n$ except for the appropriate change of range.
Thus the result of computing $\Xi^{V,\lambda} * \Xi^{V,\lambda}$
is:
\ben \label{epp4}
\langle \Xi^{V,\lambda} * \Xi^{V,\lambda}|\phi \rangle &=&
\lim_{m,n\to\infty}  \Biggl\langle 
\exp\Big(-\lambda \int_{ {\pi\over 4}}^{({m\over 2}
-{1\over
4})\pi} V(x_{m+n-1})
dx_{m+n-1}\Big) 
\\
&& \exp\Big( - \lambda \int_{ 
({m\over 2}-{1\over 4})\pi}^{ ({n\over 2} -{1\over
4})\pi + {1\over 2} (m-1)\pi} V(x_{m+n-1})
dx_{m+n-1}\Big)
f\circ
\phi(0)\Biggr\rangle_{\wh C_{m+n-1}} \, .  \nonumber
\een
This can be rewritten as
\be \label{epp5}
\langle \Xi^{V,\lambda} * \Xi^{V,\lambda}|\phi \rangle
=\lim_{m,n\to\infty}  \Biggl\langle 
\exp\Big(\hskip-4pt -\lambda \int_{{\pi\over 4}}^{{m+n-1\over
2}\pi -{\pi\over 4}} V(x_{m+n-1})
dx_{m+n-1}\Big)
f\circ
\phi(0)\Biggr\rangle_{\wh C_{m+n-1}}  .  
\ee
This is precisely eq.\refb{epp2} with $n$ replaced by $m+n-1$ (which is
taken to infinity). This establishes the required result \refb{epp3}.

In composing the integrals in passing to \refb{epp5} the
insertions of $V$ do coincide at
one point, 
namely at $x_{m+n-1}= ({m\over 2} - {1\over 4})\pi$. This will cause
additional divergences even if the exponentials appearing in the
definition of each $\Xi^{V,\lambda}$ are ultraviolet regularized.
This divergence can 
be regulated as in 
subsection \ref{sn1}  
by defining
$\Xi^{V,\lambda}$ 
through eq.\refb{epp2}, with the 
$x_n$ integral
running from ${\pi\over 4} + \epsilon$ to $({n\over 2} -{1\over
4})\pi -\epsilon$. Doing so, we
miss in the exponent of eq.\refb{epp5}
an integral of $\lambda V$ over the range $(({m\over 2} - {1\over 4})\pi
-\epsilon)\le x_{m+n-1}\le (({m\over 2} - {1\over 4})\pi
+\epsilon)$ compared to the expression for
$\langle\Xi^{V,\lambda}|\phi\rangle$.
In the $m,n\to\infty$ limit this region moves off to infinity, and as
a result contributes
an $\epsilon$ dependent multiplicative factor
to the correlation function given by the expectation value of the
missing operator in the deformed theory.\footnote{We are assuming
here
that the correlation function of the missing operator in the deformed
theory satisfies cluster property.} This is the analog of the factor
of  $(2\epsilon)^{-2h}$ which arose in the analysis of 
subsection \ref{sn1},
and can be absorbed in the definition of $\Xi^{V,\lambda}$ to ensure
that it squares to itself under $*$-multiplication.

Since the operator $V$ is in the matter sector, 
$|\Xi^{V,\lambda}\rangle$ has the factorized form
\be \label{epfacg}
|\Xi^{V,\lambda}\rangle=|\Xi_g\rangle \otimes |\Xi^{V,\lambda}_m
\rangle\, ,
\ee
where $\Xi_g$ is independent of $V$ and $\lambda$. 
As before, with $\Xi_g*^g\Xi_g=\Xi_g$ we  ensure that
\be \label{eprodma} 
\Xi^{V,\lambda}_m *^m \Xi^{V,\lambda}_m = \Xi^{V,\lambda}_m\, .
\ee
Thus we can now construct new D-brane solutions by taking the product
$|\Psi_g\rangle\otimes |\Xi^{V,\lambda}_m\rangle$, where
$|\Psi_g\rangle$ 
is the 
universal ghost state that appears in the 
D25-brane solution. 

\subsection{Computation of the tension}  \label{sn0} 

We would now like to compute the tension associated with this solution.
This is proportional to $\langle
\Xi_m^{V,\lambda}|\Xi_m^{V,\lambda}\rangle_m$.\footnote{Since this
computation involves matter sector fields only, there may be overall
constant factors appearing at various stages due to non-zero central
charge of the bulk matter theory. These 
are independent of $V$ and $\lambda$ and cancel when we compute
the ratio of tensions.}  
Computation of this inner product is again simple
in the $\wh z_n$ coordinates 
and the relevant geometry was discussed
above \refb{exy5}. We therefore find
\be \label{epp6}
\langle \Xi_m^{V,\lambda}|\Xi_m^{V,\lambda}\rangle_m =
\lim_{m,n\to\infty}  \Bigl\langle 
\exp\Big(-\lambda \int_{ {\pi\over 4}}^{{m+n-1\over
2}\pi -{\pi\over 4}} V(x_{m+n-2})
dx_{m+n-2}\Big)
\Bigr\rangle_{\wt C_{m+n-2}} \, , \nonumber \\
\ee
where $x_{m+n-2}=\Re(z_{m+n-2})$. 
If we do a more careful  
analysis using the regularized form of
$\Xi^{V,\lambda}$, there will be two missing regions, each of width
$2\epsilon$, in the exponent of eq.\refb{epp6}. 
These regions are infinite
distance apart in the $m,n\to\infty$ limit and each of these regions give
rise to an extra
multiplicative factor, equal to the one that appeared in the computation
of $\Xi^{V,\lambda}*\Xi^{V,\lambda}$, in eq.\refb{epp6}. These exactly
cancel the
$\epsilon$-dependent normalization of $\Xi^{V,\lambda}$ 
required to ensure
that the regularized $\Xi^{V,\lambda}$ squares to itself under
$*$-multiplication. This is analogous to the corresponding
phenomenon in section \ref{sn1}.

We now define a rescaled coordinate $u$ as 
$u = 4(z_{m+n-2}-{\pi\over 4})/(m+n-2)$   
so that $\Re(u)$ 
ranges from 0 to $2\pi$. Thus in the
$u$ coordinate we have a semiinfinite cylinder $C$ of circumference
$2\pi$.
Writing $u=i\rho+\theta$, and 
taking into account the conformal transformation of the vertex
operator $V$ under this scale transformation, 
we get:
\be \label{epp7}
\langle \Xi_m^{V,\lambda}|\Xi_m^{V,\lambda}\rangle =
\lim_{m,n\to\infty}  \Bigl\langle
\exp\Big(-\lambda_R \int_{\theta=0}^{2\pi}
d \theta V_R(\theta) \Big) \Bigr\rangle_{C}\, ,
\ee
where $\lambda_R\int V_R$ denotes the operator to which the 
perturbation $\lambda \int V$ flows under the rescaling by 
$(m+n-2)/4$. 
In particular, 
if $V$ has dimension $h$, it 
does not 
mix with other operators, and its conformal dimension remains unchanged 
under the renormalization group flow, then we have
\be \label{epp7a}
\lambda_R \int_{\theta=0}^{2\pi} d\theta V_R(\theta) =  \lambda ({1\over 
4}(m+n-2))^{1-h}\int_{\theta=0}^{2\pi}
d \theta V(\theta)\, .
\ee
This semiinfinite cylinder in the $u$ coordinate is
nothing but a unit disk $D_U$ with $\theta$ labeling the angular
parameter along the boundary of the disk, and $e^{-\rho}$ 
labeling the radial coordinate. Thus what \refb{epp7} represents is 
the partition
function on a unit disk, with the perturbation 
$\lambda_R \int V_R(\theta) d\theta$
added at the boundary!
Notice now that if $V$ is 
exactly
marginal, {\it i.e.} if $h=1$ to all orders, 
then the
perturbation is simply $-\lambda \int_{0}^{2\pi}
d \theta V(\theta) $, whereas if 
$V$ is a relevant deformation then $h<1$
and 
in the limit $m,n\to\infty$, $\lambda_R\int d\theta V_R(\theta)$ 
approaches its infrared fixed point $\lambda_{IR} \int d\theta 
V_{IR}(\theta)$.\footnote{We shall not consider 
irrelevant perturbations as
it is not clear how to tame the resulting ultraviolet divergences. 
Since they flow to zero in the IR, such  perturbations are not expected
to 
give rise to new solutions.}  
Thus the net result is that 
$\langle \Xi_m^{V,\lambda}|\Xi_m^{V,\lambda}\rangle_m$ represents the
partition
function on the unit disk of the BCFT to which 
the theory flows in the
infrared! As discussed below equation \refb{exy6} this 
is indeed the tension of the D-brane associated to this BCFT. 
Thus we
conclude that $|\Psi_g\rangle\otimes |\Xi_m^{V,\lambda}\rangle$ 
gives the D-brane solution
corresponding to the 
BCFT to which the theory would flow in the infrared 
if we added to the action 
the boundary perturbation  proportional to $\int V(\theta)
d\theta$. 
We emphasize 
that the string field
$|\Psi_g \rangle \otimes |\Xi_m^{V, \lambda}\rangle$
belongs to the state 
space of BCFT$_0$.

In particular if we take $V$ to be the identity operator  $I$, 
we see from eq.\refb{epp2} that 
\be \label{epp22a}
\langle \Xi^{I,\lambda}| \phi\rangle = \lim_{n\to\infty} 
\Bigl\langle 
\exp(-\lambda \int_{ \pi/4}^{({n\over 2} -{1\over 4})\pi}
dx_n) f\circ \phi(0)\Bigr\rangle_{\wh C_n}\, , \qquad
\forall
|\phi\rangle\in\HH\, .
\ee
For $\lambda>0$, this vanishes due to the infinite range of integration
over $x_n$. Thus we get $|\Xi^{I,\lambda}\rangle=0$ and hence
$|\Xi_m^{I,\lambda}\rangle=0$. This is consistent
with the fact that the trivial solution $\Psi=0$ represents the
tachyon vacuum
configuration, and that in the boundary string field theory formalism,
perturbation by the identity operator takes 
the unstable D-brane 
to the tachyon vacuum.

Note, however, 
that the solution \refb{epp2} does seem to depend
on $\lambda$ for more general relevant perturbations. Since different
values of $\lambda$ correspond to the same tension of the final brane, we
expect that they represent gauge equivalent solutions. Thus the
parameter $\lambda$
is analogous to the redundant parameter $b$ labeling the lower
dimensional D-$p$-brane solutions considered in
ref.~\cite{0102112}.\footnote{This was suggested to us by
Witten~\cite{PRIVATE}.}

\subsection{Small deformations of the sliver}  
\label{s4.2}

{}From the analysis of subsection \ref{s4a} it is clear that
when the operator $V$ is relevant, the 
state $|\Xi^{V,\lambda}\rangle$ 
corresponds to a big change in field 
configuration since it gives rise to
a totally new D-brane solution 
corresponding to the BCFT to which
the theory flows 
in the infrared,
whereas if $V$ is exactly 
marginal, then for small
$\lambda$, $|\Xi^{V,\lambda}\rangle$ is a 
solution `close to' the original
solution $\Xi$.\footnote{Even though  
$\langle\Xi_m|\Xi^{V,\lambda}_m\rangle_m$ vanishes even for small 
$\lambda$, we treat the solutions as `close' 
in the sense that the BCFT's 
corresponding to $\lambda=0$ and $\lambda$ small have correlation 
functions which differ from each other by small amount.}
Thus we can 
define a small 
deformation $\del\Xi^V$ around the solution $\Xi$ through the relation:
\be \label{eqq1}  
\Xi^{V, \lambda} = \Xi + \lambda \del\Xi^V + O(\lambda^2) \, ,
\ee
and using eqs.\refb{epp2} and  \refb{epp22} we find 
\ben \label{eglpre}
\langle \del\Xi^V| \phi\rangle &=& -\lim_{n\to\infty} \Bigl\langle 
\int_{{\pi\over 4}}^{({n\over 2} -{1\over 4})\pi}
V(x_n) dx_n  f\circ
\phi(0)\Bigr\rangle_{\wh C_n} \nonumber \\
&=& -\Bigl\langle 
\int_{|x'|>{\pi\over 4}} V(x') dx'  f\circ
\phi(0)\Bigr\rangle_{D_H} \, ,\qquad \forall
|\phi\rangle\in\HH\, .
\een

By expanding eq.\refb{epp3} in powers of $\lambda$, it follows
that $\del\Xi^V$ satisfies
\be \label{eqq3}
\Xi * \del\Xi^V + \del\Xi^V * \Xi = \del\Xi^V\, .
\ee
Thus $|\Psi_g\rangle\otimes |\del\Xi^V_m\rangle$ will describe small
deformation of the string field theory solution describing the
D-brane corresponding to BCFT$_0$.  
One can
also give a direct proof of eq.\refb{eqq3} by using the $*$-product rules
and the
gluing relations \refb{ecor4} as follows. For this we shall 
represent both
$\del\Xi^V$ through the first of eq.\refb{eglpre} and $\Xi$ by a
similar formula without the $\int V(x_n) dx_n$ insertion for a fixed
finite $n$ and then take $n\to\infty$ limit at the end of the 
calculation.
Using the by now standard procedure we get:
\ben \label{eglue}
\langle \del\Xi^V*\Xi|\phi\rangle &=& 
-\lim_{n \to \infty} \,\Bigl\langle \,\int_{{\pi\over  4}}^{
\pi ({n\over 2}-{1\over 4})} d x_{2n-1}
V(x_{2n-1}) \; f\circ \phi (0) \Bigr\rangle_{\wh C_{2n-1}}
\nonumber \\
\langle \Xi * \del\Xi^V|\phi\rangle &=&
-\lim_{n \to \infty} \,\Bigl\langle \,\int_{\pi ({n\over 2}-{1\over
4})}^{
\pi (n-{3\over 4})} d
x_{2n-1} \, V(x_{2n-1}) \;  f\circ \phi (0) 
\Bigr\rangle_{\wh C_{2n-1}} 
\,.
\een
Thus we have:
\be \label{eglue2}
\langle \del\Xi^V*\Xi|\phi\rangle + \langle \Xi *
\del\Xi^V|\phi\rangle = 
-\lim_{n \to \infty} \,\Bigl\langle \,\int_{{\pi\over 4}}^{
\pi (n-{3\over 4})} d
x_{2n-1} \, V(x_{2n-1}) \;  f\circ \phi (0) \Bigr\rangle_{\wh C_{2n-1}}
\,.
\ee
Introducing $m=2n-1$ we can rewrite this as 
\be \label{eglue3}
\langle \del\Xi^V*\Xi|\phi\rangle + \langle \Xi *
\del\Xi^V|\phi\rangle =
-\lim_{m \to \infty} \,\Bigl\langle \,\int_{{\pi\over  4}}^{
\pi ({m\over 2}-{1\over 4})} d x_m
\, V(x_m) \; f \circ \phi (0) \Bigr\rangle_{\wh C_m} 
\,.
\ee
This is precisely the expression \refb{eglpre} for $\langle
\del\Xi^V|\phi\rangle$. Thus we see that $\del\Xi^V$ satisfies
equation \refb{eqq3}.

For later use 
we need to use  a regularized expression for $\del\Xi^V$ by restricting
the integration range in \refb{eglpre} to be $|x'|\ge {\pi\over
4}+\epsilon$, or equivalently, ${\pi\over 
4}+\epsilon \le x_n \le ({n\over 2}-{1\over 4})\pi -\epsilon$. When we
compute the analog of \refb{eglue2} with this regularized expression for
$\del\Xi^V$, and compare this with the regularized version of
\refb{eglpre}, we find that we miss an integral 
\be \label{emissing}
-\int_{({n\over 2}-{1\over 4})\pi -\epsilon}^{({n\over 2}-{1\over 4})\pi
+\epsilon} dx_{2n-1} \Bigl\langle\, V(x_{2n-1}) f\circ \phi(0)
\Bigr\rangle_{\wh C_{2n-1}}\, . 
\ee
In the $n\to\infty$ limit the integration region 
moves off to infinity and
the correlator above vanishes as long as $V$ has dimension $>0$. Thus the
regularized $\del\Xi^V$ still satisfies eq.\refb{eqq3}. 
In the next subsection we shall interpret $\del\Xi^V$ as an
appropriate covariant derivative of the surface state along 
a  marginal
direction in the space of boundary conformal field theories.

\subsection{Background independence and theory-space connections} 
 \label{ss31}

The string field theory action around the tachyon vacuum
enjoys a fundamental property: it has manifest 
background independence with respect to the open string moduli.
This observation will allow us to recover the construction
of infinitesimal deformations of 
classical solutions
from the point of view of deformations 
along marginal directions in the open string theory 
moduli space.
To explain these ideas, we need to recall some notions about
connections in theory space. Our discussion will follow
\cite{RSZ, SZ}, with some obvious modifications
needed for open strings.

Let us fix the CFT in the bulk (the closed string moduli)
and for this given bulk CFT, let us
consider the space of BCFT's, labeled
by some continuous parameters $\{x^s\}$.  
Each point $\{x^s\}$ specifies
a BCFT, with its own state space $\HH_x$.
We have the structure of a vector bundle,
where the base  is theory space (parameterized
by $\{x^s\}$), and the fiber at each point $x$ is $\HH_x$.
Thus we can introduce connections on this vector bundle. 
While there are many possible
connections, a natural choice for open string field theory 
is the canonical connection $\widehat \Gamma$ \cite{RSZ}. 
On surface states, the associated
covariant derivative ${\widehat D}_s $ 
acts as the integration of the 
exactly marginal perturbation $\OO_s$
over the boundary of the Riemann surface, 
excluding from the integration region
the segments around the punctures where local coordinates are defined:
\be \label{ekk1}
{\widehat D}_s \, \langle \Sigma| = - \int_{\partial\Sigma - \cup_i \,
D_i}  d z' \; \langle \Sigma; z' \,|\OO_s \rangle\, . 
\ee
Here $\Sigma$ is a disk with $n$ punctures
on its boundary $\partial \Sigma$ 
with some global coordinate $z$. 
The regions 
$D_i \subset \p\Sigma$ are the images of the diameters of local
coordinates
half-disks $\{|\lo_i| \leq 1, \Im(\lo) =0 \}$. 
The surface state $\langle \Sigma; z'|$
is the $n+1$-punctured disk obtained by introducing
an extra puncture on $\Sigma$ at the boundary point $z=z'$, 
with local coordinate $\lo= z - z'$. 
Finally $\OO_s$ is 
an exactly marginal operator 
of the boundary CFT  
inserted at this new puncture.

The reflector state\footnote{The reflector ${}_{12}\langle R|$ is
the two-punctured disk with punctures at $z=0$ and
$z=\infty$ in UHP coordinates, with local coordinates
$\lo_1=z$ around $z=0$ and $\lo_2=-1/z$ 
around $z =\infty$.
The reflector provides the definition of the BPZ inner 
product: $\langle \psi | \phi  \rangle \equiv  
{}_{12}  \langle R| \; \psi \rangle_1 \,|\phi \rangle_2$. }
 ${}_{12}\langle R|$ and the cubic vertex
$ {}_{123}\langle V_3 |$ are covariantly constant with respect
to this connection:
\be \label{covconst}
\widehat D_s\, {}_{12}\langle R| =0 \,, \qquad \widehat D_s\, 
{}_{123}\langle V_3 | =0\,.
\ee
This follows immediately from the fact
that the local coordinate patches 
cover completely the two and three punctured disks associated
to ${}_{12} \langle R|$ and ${}_{123} \langle V_3|$,
so that in both cases $\p\Sigma - \cup_i D_i$ 
is the empty set.
We can rephrase (\ref{covconst}) by saying
that $\widehat D_s$ acts as a derivation of the $*$ product:
\be \label{Dderiv}
\widehat D_s (\phi_1 * \phi_2) = \widehat D_s \phi_1 * \phi_2
+ \phi_1 * \widehat D_s \phi_2 \,,  
\ee
where $\phi_1$ and $\phi_2$ are two sections of the vector bundle.

At each point $x$ in theory space, exactly 
marginal
deformations are represented by 
dimension one primaries $\OO_s$ of the 
{\it matter} BCFT. 
Thus for each exactly 
marginal direction $s$,
the covariant derivative $\widehat D_s$ 
does not involve any ghosts and commutes
with the purely ghost kinetic operator $\QQ$ of vacuum
string field theory, 
\be \label{[DQ]}
[\widehat D_s, \QQ] = 0 \,.
 \ee
The string field action \refb{eo1} is 
naturally a 
function on the vector bundle. 
$\SS(\Psi)$ is indeed 
$\SS(\psi^i, x)$ where $\psi^i$ are components along the
vectors in $\HH_x$. It can be written as
\be \label{eo1new}
\SS (\Psi, x) \equiv \,-\, {1\over g_0^2}\,\,\bigg[\, {1\over 2}
\langle R |
 \Psi\rangle \QQ|\Psi\rangle
 + {1\over 3}\langle V_3| \Psi\rangle|\Psi\rangle
|\Psi\rangle
 \bigg] \,.
\ee
It follows from the definition of a covariant derivative
of functions on a vector bundle (such derivative measures
how the function changes as we move on the base, and on
the fiber by parallel transport) and relations 
\refb{covconst} and 
\refb{[DQ]} that
\be
\label{bind}
\widehat D_s\, \SS = 0 \,.
\ee
This is the statement of manifest background independence
of the vacuum string field theory: there is a (canonical)
connection $\wh\Gamma$ so that the action is covariantly constant.
This means 
that  vacuum string field theory is independent of the choice of
the  reference boundary conformal field theory BCFT$_0$ $-$
string field actions using nearby BCFT$'$s are identical when the
string fields are related by an 
homogeneous field redefinition generated by parallel transport with 
the connection $\wh\Gamma$. This is simpler than for conventional string 
field theory~\cite{SZ}, where the covariant derivative of the
action is nonvanishing and  the requisite
field redefinition includes a constant shift and other terms in 
addition to parallel transport.

Consider now a solution 
$\Psi^x=\Psi_g\otimes\Psi^x_m$ of  
the vacuum  
string field theory equations corresponding to a certain
BCFT background labeled by $x$, 
{\it expressed as an element of
$\HH_x$.} $|\Psi^x_m\rangle$ is simply
the
matter part of the sliver state $|\Xi^x_m\rangle$ associated with this
conformal field theory.
Thus we have
\be \label{psixeq}
\QQ \Psi^x + \Psi^x * \Psi^x = 0\, , \qquad \Psi^x_m *^m \Psi^x_m =
\Psi^x_m\, .
\ee
Since $\Psi^x$ is defined as an element of $\HH_x$ for every $x$, we have
a section of the vector bundle introduced earlier. Using the fact that
$\widehat D_s |\Psi_g\rangle=0$, we have
\be \label{firstx}
\widehat D_s |\Psi^x\rangle = |\Psi_g\rangle\otimes \widehat D_s 
| \Xi^x_m\rangle\, .
\ee 
Since $|\Xi^x_m\rangle$ is a surface state, its covariant derivative is
given by eq.\refb{ekk1}. Comparing this with \refb{eglpre} we see that
this is just $|\del\Psi^{\OO_s}\rangle$.  Thus we have
\be \label{secondx}
\widehat D_s |\Psi^x\rangle = |\Psi_g\rangle\otimes |\del\Psi^{\OO_s}
\rangle\, .
\ee
The small deformation  
of the sliver solution considered in the
previous subsection is simply the covariant derivative of $\Psi^x$ with
the
connection $\widehat\Gamma$.
Indeed, applying $\widehat D_s$ 
on the second of  
eq.\refb{psixeq}, and
using
eqs.\refb{Dderiv}, 
one gets
\be \label{deformeq} 
(\widehat D_s\Psi^x_m) *^m \Psi^x_m + \Psi^x_m*^m(\widehat 
D_s\Psi^x_m)=
\widehat D_s \Psi^x_m\, . \nonumber
\ee
This shows that $\widehat D_s \Psi^x$ satisfies the small fluctuation
equation \refb{eqq3}. 

Although we have been careful in phrasing the discussion
in terms of {\it exactly} marginal deformations,
as only for those it makes sense to talk about connections
in theory space, for {\it infinitesimal} deformations 
we can nevertheless consider the generalization  
that $\OO_s$ is a generic dimension one matter primary.
The operator $\widehat D_s$ is still well-defined
and obeys (\ref{Dderiv})   and (\ref{[DQ]}). So 
we obtain a solution $\widehat D_s \Psi^x$ of the 
small fluctuation equations for each dimension
one matter primary. These states 
are  promising candidates
for the spectrum of physical excitations around
the D-brane background. We shall discuss this  
in detail in section 5.

It is intriguing that   
some of 
the previous analysis appears to apply formally 
to closed string deformations.
A closed string deformation of a surface state
is obtained  by integrating a closed
string marginal operator $\OO_{\bar s}$ 
over  the {\it bulk} of the surface state, 
leaving out the local coordinate regions. 
It would seem that such deformation of the sliver would satisfy
the small fluctuation equation. 
In this case, however,
this integral 
is plagued with ultraviolet divergences from the region of
integration where the operator $\OO_{\bar s}$ comes close to the boundary
of the disk, and it is not {\it a priori} clear how 
such divergences should
be regulated. 
This deformation $-$ if it can be defined $-$ might 
just be the open string by-product of a shift in the closed string
background, as is possibly the case in open/closed 
string field theory~\cite{Zwiebach:1998fe}.
But perhaps it represents some true encoding of
closed string physics into the open string state space.
It is worth noting that the sliver, as opposed
to the identity string field used earlier for
encoding closed string deformations in open string
theory \cite{Strom}, has plenty of bulk to insert
closed string states.
Clearly this issue deserves
further investigation.

\sectiono{Physical states around D-brane
backgrounds} \label{s6}

Given a candidate solution representing a 
specific D-brane, one would like
to find the spectrum of physical states around the D-brane background and
compare this with the known spectrum of physical open string states in
that particular D-brane background. Let $\Psi_0 
= \Psi_g \otimes \Psi_m$  be a
solution representing a particular D-brane in the string field theory
around
the tachyon vacuum \cite{0102112}. 
After shifting the string field $\Psi = \Psi_0 + \Phi$, the action takes
the form:
\be \label{eaction}
\SS (\Phi) \equiv \,-\, {1\over g_0^2}\,\,\bigg[\, {1\over 2} \langle
\,\Phi \, ,
 \, \QQ_0\, \Phi
\rangle + {1\over 3}\langle \,\Phi \, , \, \Phi *
\Phi \rangle \bigg] \,,
\ee
where the action of the BRST operator $\QQ_0$ in the new background
on any state $|A\rangle$ in $\HH$ is given as 
\be
\QQ_0\, A = \QQ A + \Psi_0 * A - (-)^A A* \Psi_0\, .
\ee 
Around the new background
the physical state 
condition, corresponding to the linearized
equations of
motion, is then
\be
\label{psc}
\QQ_0 \Phi = \QQ \Phi + \Psi_0 * \Phi +  \Phi * \Psi_0  = 0\, ,
\ee
while the linearized gauge
symmetries are given as
\be \label{gauge}  
\delta_\Lambda \Phi = \QQ_0\Lambda = \QQ \Lambda + \Psi_0 * \Lambda -
\Lambda* 
\Psi_0 \,.
\ee
Physical states around $\Psi_0$ 
are identified with the cohomology classes of $\QQ_0$.

Clearly a full solution of this problem requires the knowledge of $\QQ$,
and the solution $\Psi_0$ including its ghost part, which are not
available to us at present.
In this section we shall try to get some insight into this problem.

\subsection{Factorization ansatz for fluctuations} \label{s4.1}

Clearly there are two parts to the problem. First we need to find
solutions to the linearized equations of motion \refb{psc} and then we
need to determine which of these are 
related by gauge transformations. For
the first problem, we could look at a subset of field configurations of
the form:
\be \label{factorization}
\Phi =   \Psi_g \otimes \delta\Psi_m  \, ,
\ee
{\it where  $\delta\Psi_m$ is a matter state}. 
Replacing 
\refb{factorization} into
\refb{psc}
and using eq.\refb{eo4}  
we find the condition
\be
\label{fpsc}
\Psi_m *^m \delta \Psi_m
+ \delta \Psi_m*^m \Psi_m = \delta \Psi_m \,.
\ee
Note that this condition does not involve the operator
$\QQ$.
Clearly not every solution of \refb{psc} is of the form
\refb{factorization}. However it is
conceivable that every solution of \refb{psc} can be brought to this form
using a gauge transformation \refb{gauge}, {\it i.e.} there is a
representative element of the form \refb{factorization} in every
cohomology class of $\QQ_0$.  

It is possible to find solutions to eq.\refb{fpsc}. 
For definiteness we shall take $\Psi_m$ to be $\Xi_m$, 
$-$ the matter part of the sliver state associated with BCFT$_0$, $-$
and 
suppose that it 
represents a particular D-brane in a moduli space that
includes some marginal directions. 
As discussed in section \ref{s3},
the various points in this moduli space
can be obtained as 
deformed sliver solutions. 
Defining $\del\Xi^V_m$ as the matter part of $\del\Xi^V$ defined in
section \ref{s4.2} and using eq.\refb{eqq3}
we see that 
$\delta\Psi_m\propto\del\Xi^V_m$  
satisfies
eq.\refb{fpsc}.
Although the argument given above uses an exactly marginal deformation,
the explicit analysis of section \ref{s4.2}, showing that $\del\Xi^V$
satisfied eq.\refb{eqq3}, goes through if we choose $V$ to be an
arbitrary dimension one primary operator. As a result
$\delta\Psi_m\propto\del\Xi^V_m$ 
still 
satisfies eq.\refb{fpsc}. Thus we seem to
have gotten a solution of the linearized equations of motion for every
dimension one matter primary operator in BCFT$_0$.

While this is encouraging,  
there are some subtleties which we now discuss.
First of all,  we compute the inner product 
$\langle\del\Xi^V_m|\del\Xi^V_m\rangle$
as in  eq.\refb{epp6}. This is given by:
\ben \label{einner2}
\langle \del\Xi_m^V|\del\Xi_m^V\rangle &=&  
\lim_{m,n\to\infty}  \Big\langle 
\Big(\int_{{\pi\over 4}}^{({m\over 2} -{1\over
4})\pi} V(x_{m+n-2})
dx_{m+n-2}\Big) \nonumber \\
&&
\Big( \int_{({m\over 2} -{1\over 
4})\pi}^{{m+n-1\over
2}\pi -{\pi\over 4}} V(x_{m+n-2})
dx_{m+n-2}\Big)
\Big\rangle_{\wt C_{m+n-2}} \, . 
\een
The correlation functions are computed in the matter sector only.
In the $m,n\to\infty$ limit the range of integration over $x_{m+n-2}$ is
infinite, but this is easily avoided by going to a new coordinate system
$u\equiv i\rho+\theta=4(z_{m+n-2}  
-\pi/4)/(m+n-2)$ which maps $\wt
C_{m+n-2}$ to a seminfinite cylinder $C$ of circumference $2\pi$.
Dimension one primary
fields will not pick up any factor under this rescaling, and we get
\be \label{einner1}
\langle \del\Xi^V_m|\del\Xi^V_m\rangle =
\Bigl\langle 
(\int_{0}^{\pi} V(\theta)
d\theta)
( \int_{\pi}^{2\pi} 
V(\theta') d\theta')
\Bigr\rangle_{C} \, . \nonumber \\
\ee
Note now that the range of integration over $\theta$ and $\theta'$
coincide
at two points, $\theta=0$ and $\theta=\pi$. Since $\langle V(\theta)
V(\theta')\rangle \sim |\theta-\theta'|^{-2}$ 
for small $ |\theta-\theta'|$,
this gives logarithmically
divergent integrals. Thus the BPZ norm of $\del\Xi^V_m$ in the matter
sector is logarithmically 
divergent.  
This of
course is related to 
the ultraviolet divergence discussed in section \ref{s3}
and
one could try to regularize this by taking the limit of $x'$ integration
in eq.\refb{eglpre} to be in the range $|x'|\ge{\pi\over 4}+\epsilon$. As
discussed 
in section \ref{s4.2},  
$\del\Xi^V$ satisfies eq.\refb{eqq3} even when
$\epsilon$ is finite. 
However unlike 
$\Xi_m^{V,\lambda}$, which has finite norm for finite $\epsilon$ (when
measured with respect to the
norm of $\Xi_m$), $\Delta\Xi^V_m$ has infinite norm even for finite
$\epsilon$. To see this we note that the rescaling needed to go from the
$x_{m+n-2}$ coordinate to the $\theta$ coordinate shrinks a region of
finite width to zero width. More precisely, 
with the regulator $\epsilon$
in place and after the rescaling, we get
\be \label{einner1a}
\langle \del\Xi^V_m|\del\Xi^V_m\rangle = \lim_{m,n\to\infty} 
\Biggl\langle 
(\int_{4\epsilon/(m+n-2)}^{\pi-4\epsilon/(m+n-2)} V(\theta)
d\theta)
( \int_{\pi+4\epsilon/(m+n-2)}^{2\pi-4\epsilon/(m+n-2)} 
V(\theta') d\theta')
\Biggr\rangle_{C} \, . \nonumber \\
\ee
Thus even for finite $\epsilon$ the integral has logarithmic divergences
in the $m,n\to\infty$ limit, since the 
integration limits  
approach each other. 
This divergence is actually needed for consistency. 
Indeed, note that 
for a marginal deformation $\Xi^{V,\lambda}_m$ can be identified with
$\Xi'_m$ associated with a deformed BCFT$'$ discussed in section
\ref{sn1}. The norm of $\Xi'_m$, being equal to 
$(2\epsilon)^{-4h}$ times  
the partition function of BCFT$'$ on the cylinder, approaches that of
$\Xi_m$ for small $\lambda$ since as $\lambda\to 0$, $h\to 0$ and
the partition functions of BCFT$'$ and BCFT$_0$ are equal.
The orthonormality of $\Xi_m$ and $\Xi^{V,\lambda}_m=\Xi'_m$ (section
\refb{sn1}) then leads to the result that the 
norm of  $(\Xi^{V,\lambda}_m
-\Xi_m)$ is twice the norm of $\Xi_m$ in the $\lambda\to 0$ limit. 
Given that $\Xi^{V,\lambda}_m = \Xi_m
+ \lambda\Delta\Xi^V_m + \OO (\lambda^2)$,  we reach the conclusion that
$\lambda^2$ times the norm of $\Delta\Xi^V_m$ is twice the norm of
$\Xi_m$ 
in the small $\lambda$ limit.
If this has to be true for arbitrarily small
$\lambda$, then  $\Delta\Xi^V_m$ must have infinite norm.

This divergence is not necessarily
problematic, 
since it is not clear {\it a priori} that 
having
finite norm in the matter sector is a requirement on physical
deformations. 
As we have seen in section \refb{sn1}, the deformed solution 
$\Xi'_m$   
does have finite norm, 
even if $\Delta\Xi^V_m$ has divergent
norm.  We shall also argue later that having 
a logarithmically divergent norm of this kind may 
be a necessary condition for a physical deformation not to be pure
gauge. 
{}From a physical viewpoint, since we have seen that for exactly
marginal 
$V$, $\Delta\Xi^V$ represents some small deformation of BCFT$_0$, we are 
led to believe that such deformations must be allowed in the computation 
of the spectrum.

The second subtlety arises  because 
the manipulations in
section
\ref{s4.2} leading to eq.\refb{eqq3} actually hold  
without any
constraint on $V$ as the gluing in the $\wh z_n$ 
coordinate system does not
require us to make any conformal transformations at all. 
So one might ask:
why can't we use the matter part of these states to find new solutions of
eq.\refb{fpsc}? The answer to this question may again be hidden in the
normalization of the state. For this, let us write down the norm of
$\del\Xi^V_m$ for an operator $V$ of arbitrary dimension $h$:
\ben \label{einner1b}
\langle \del\Xi^V_m|\del\Xi^V_m\rangle &=& \lim_{m,n\to\infty}  
\Big[\Big({m+n-2\over 4}\Big)^{2(1-h)}
\Big\langle 
\Big(\int_{4\epsilon/(m+n-2)}^{\pi-4\epsilon/(m+n-2)}
V(\theta)
d\theta\Big) \nonumber \\  
&&
\Big(
\int_{\pi+4\epsilon/(m+n-2)}^{2\pi-4\epsilon/(m+n-2)}
V(\theta') d\theta'\Big)
\Big\rangle_{C}\Big] \, . 
\een
Note the prefactor coming from the conformal transformation of the vertex
operators when we make the final conversion from $x_{m+n-2}$ to $\theta$
coordinate. 
For $h<1$ there is no short distance divergence from the
limits of integration
of $\theta$, $\theta'$, but the contribution from the 
finite $\theta$ region 
has power law divergence due to the infinite prefactor in the
$m,n\to\infty$ limit. Thus the divergence in the norm of such states is
worse than that of a marginal deformation,  and clearly originates from
the large distance region (in the 
original $x_{m+n-2}$ coordinate system.)
In fact this divergence is precisely the reason
why a small perturbation by $V$ drives the system all the way to the
infrared fixed point. Thus we can conclude
with reasonable degree
of confidence that for $h<1$ the $\del\Xi^V$ does not represent small
deformation. On the other hand for $h>1$ the 
integral is finite for finite
$\epsilon$ as the short distance singularity coming from the region of
integration $\theta\simeq\theta'$ at 0 or $\pi$ cancel the conformal
factor exactly. Thus $\del\Xi^V$ has finite norm. 
This seems to make
irrelevant perturbations better behaved.
As we shall see below, however, these are most likely pure gauge
deformations, precisely because 
they have finite norm.

\subsection{Factorization ansatz for gauge transformations}
\label{ssgauge}

Not all solutions
of
\refb{fpsc} represent inequivalent physical states, since we still need
to consider the linearized gauge invariances (\ref{gauge}).
A gauge transformation (\ref{gauge})
with generic parameter $\Lambda$ will
not respect the form (\ref{factorization})
of the fluctuations. While there may be more
general gauge parameters respecting  \refb{factorization}
we  now restrict ourselves to 
factored gauge parameters
$\Lambda = \Lambda_g \otimes \Lambda_m$,
where $\Lambda_m$ is a matter state and 
$\Lambda_g$ is a ghost state satisfying 
\be \label{elamg}
\QQ \Lambda_g = 0, \qquad
\Lambda_g *^g \Psi_g = \Psi_g*^g\Lambda_g  =
\Psi_g\, .
\ee
For such a $\Lambda_g$, (\ref{gauge}) gives
\ben
\label{4p7}
\delta_\Lambda (\Phi) & =&
\QQ \Lambda_g \otimes \Lambda_m 
+ (\Psi_g *^g\Lambda_g )\otimes (\Xi_m *^m \Lambda_m - 
\Lambda_m *^m \Xi_m )\nonumber \\ 
\to \quad \Psi_g\otimes\delta_\Lambda \delta\Psi_m& =&
\Psi_g \otimes (\Xi_m *^m \Lambda_m - 
\Lambda_m *^m \Xi_m)  \,.
\een
It follows from this equation and from \refb{fpsc} that 
we can study physical states
around D-brane backgrounds by considering the
space of equivalence classes
$\HH_{\delta
\Psi}$ of matter string fields
$\delta \Psi_m$ which satisfy the equations:
\ben \label{physicalcond}
&&\Xi_m *^m \delta \Psi_m
+ \delta \Psi_m*^m \Xi_m = \delta \Psi_m  \\ \label{gaugeequiv}
&&\delta\Psi_m \simeq \delta \Psi_m +  
\Xi_m *^m \Lambda_m - 
\Lambda_m *^m \Xi_m \,.
\een 
We should keep in mind, however, that 
there may be gauge transformation parameters
other than those of the form $\Lambda_g\otimes\Lambda_m$
which generate factorized deformations of the form 
$\Psi_g\otimes\delta\Psi_m$. These 
will generate further equivalence 
relations between the deformations $\delta\Psi_m$.

Does a $\Lambda_g$ satisfying \refb{elamg} exist?
Let $\II_g$ denote the ghost part of the identity string state, 
normalized so that $\II_g*^g\II_g=\II_g$. If $\QQ$ 
annihilates the identity state, then
$\Lambda_g = \II_g$ 
clearly satisfies eq.\refb{elamg}. We shall now argue that 
$\Lambda_g=\II_g$ is also the most natural choice.
{}From the analysis of 
refs.~\cite{comma,BORDES,ABDUR1,ABDUR2,BORDES2,TWO,gross-taylor, kawano} 
we know that 
barring the complications of the string mid-point coordinate and/or zero 
modes, it is 
natural to think of the string field as a matrix acting on the 
half 
string state space. In particular we can think of the factored string 
field of the form \refb{eo3} as the 
direct product of two matrices, one acting on the ghost sector of the
half-string state and the other acting on the matter sector of the
half-string 
state space. Since the identity string field $\II$ can be regarded as an 
identity matrix on the half string state space, choosing 
$\Lambda=\II_g\otimes \Lambda_m$ corresponds to a matrix which is a
direct 
product of the identity matrix in the ghost sector of the half-string 
state space and 
generator of a non-trivial rotation in the matter sector of the 
half-string state space. 
This is precisely how we would represent the rotation generator on a 
product of two vector spaces if we want to construct generators that act 
on only one of the two vector spaces. Thus from this physical point of 
view choosing $\Lambda_g=\II_g$ would be most natural. This in turn,
would 
require us to choose $\QQ$ that annihilates $\II_g$. We take this as a 
strong argument in favor of choosing such $\QQ$.\footnote{The result of 
ref.\cite{0105024}, showing that around the tachyon vacuum $\II$ is a
pure 
gauge state, also requires us to choose such a $\QQ$.}
Given this choice we 
can now proceed by taking 
\refb{gaugeequiv} to 
be a valid 
gauge equivalence relation between different $\delta\Psi_m$.

Since these equations refer to matter fluctuations $\delta\Psi_m$,
without risk of confusion we shall from now on drop all
$m$ subscripts and  superscripts referring to matter and 
write these 
equations as 
\ben \label{physicalcondx}
&&\Xi * \delta \Psi
+ \delta \Psi* \Xi = \delta \Psi  \\ \label{gaugeequivx}
&&\delta\Psi \simeq \delta \Psi +  
\Xi * \Lambda - 
\Lambda * \Xi \,.
\een 
In the next two subsections we discuss  
some consequences of these equations. All states, $*$-products 
and
correlation functions in these subsections will refer to matter sector
only.

\subsection{Deformed projectors and 
rules for physical states}  
\label{s6.1.1}

The physical state conditions 
\refb{physicalcondx} obtained above  
have the interpretation of deformation equations for projectors,
with a familiar equivalence relation.  Indeed this equation
is clearly equivalent to the condition that 
$(\Xi + \delta\Psi)$ be a projector:
\be
\label{dpr}
(\Xi + \delta\Psi) * (\Xi + \delta\Psi) = (\Xi + \delta\Psi) +
\OO(\delta\Psi^2) \,.
\ee
Similarly, equation \refb{gaugeequivx} is
equivalent to 
\be
\Xi+ \delta \Psi  \simeq e^{-\Lambda} * \Xi  * e^\Lambda
+ \OO (\Lambda^2)\, ,
\ee
where in the expansion of $e^\Lambda$ 
products correspond to $*$-products.
Therefore our physical state conditions are conditions
for deformed projectors with a $U(\infty)$ local gauge
symmetry.

This leads to a puzzle.
It is simple to show formally that the above physical
state problem has no nontrivial solutions!  Indeed, 
we multiply \refb{physicalcondx} from
the left by $\Xi$ 
and use associativity and the projector condition for $\Xi$
to  obtain
\be
\label{surp}
 \Xi*\delta \Psi* \Xi = 0 \,.
\ee
It then suffices to take 
\be \label{etrivgauge}
\Lambda = \Xi * \delta \Psi- \delta\Psi*\Xi\,,
\ee
 to
verify that 
\ben
\Xi * \Lambda - 
\Lambda * \Xi  &= &\Xi*\delta \Psi  - 2 \,\Xi*\delta\Psi*\Xi 
+ \delta\Psi *\Xi\nonumber \\
&=& \Xi*\delta \Psi   + \delta\Psi *
\Xi = \delta \Psi\,,
\een
where we used associativity, equation \refb{surp} and 
\refb{physicalcondx}.  It follows that $\delta \Psi$ is
trivial.

\medskip
This clearly shows that if eq.\refb{gaugeequivx} 
is the correct 
equivalence relation, we cannot get non-trivial elements of the
cohomology using states which are completely regular. One way to avoid
this problem is to use a non-normalizable $\delta\Psi$. 
This will typically
give rise to non-normalizable $\Lambda$ of the same type  
via eq.\refb{etrivgauge}.  If  
the requirement of normalizability on string field fluctuations is
different from that on the gauge transformation parameters (which is not
improbable considering that the two are multiplied by different ghost
sector states) it is possible to have a non-normalizable $\delta\Psi$
which is an allowed deformation, while the corresponding $\Lambda$
constructed through eq.\refb{etrivgauge} is not an allowed gauge
transformation parameter. 

Indeed, this suggestion  
is physically motivated. 
We have argued earlier that despite having logarithmically divergent
norms, 
deformations $\delta\Psi\propto \del\Xi^V_m$ associated with marginal 
operators $V$ should be considered as physical deformations. It is easy
to 
verify that $\Lambda$ constructed from such a $\delta\Psi$ through 
eq.\refb{etrivgauge} also has logarithmically divergent norm.
If such 
non-normalizable $\Lambda$'s did generate gauge transformations, 
then we would be forced to 
conclude that two different BCFT's related by a marginal deformation are 
gauge equivalent.
Since this is not physically acceptable, we are led to the conclusion
that  matter sector states $\Lambda$ with logarithmically divergent norms
are not 
valid gauge transformation parameters. This principle will also explain
why  the
state associated with deformations by irrelevant operators, found at the
end of section \ref{s4.1}, are pure gauge; they have finite norm and
the
corresponding gauge transformation parameter $\Lambda$ defined through
eq.\refb{etrivgauge} also has finite norm. 
Thus such deformations can be gauged away.

While this principle leads us to identify deformations of the sliver by
dimension-one primaries as physical states, it also includes in the list
deformations by unwanted dimension-one operators $-$ the vertex
operators of null states and non-primary states. At present it is not
clear how to eliminate them and reproduce the correct spectrum of
physical states on the D-brane.\footnote{Although a null state has
vanishing two point function with itself, its two point function with a
vertex operator which is neither a primary nor a secondary, could be
non-zero.}

\sectiono{Discussion} \label{s5}  

In this concluding section we make some general remarks about the string
field theory action given in eq.\refb{eo1} and also discuss the
implication
of the results of the present paper in a general context. We end with a
discussion of some of the open problems in this subject.

\begin{itemize}

\item
Typically the formulation
of cubic open string field theory requires a choice of background,
determining the form of the quadratic term in the action. In that sense
the action \refb{eo1} represents  the choice of the
tachyon vacuum as the background around which
we expand. But 
this clearly is a
special background being the end-point of tachyon
condensation of any D-brane. The action is formally 
independent of the choice of BCFT whose basis we use to expand the
string field, as is apparent from the
fact that $\QQ$ is made purely of ghost operators, and the
$*$-product, defined through overlap conditions on string
wave-functionals, is formally independent of the choice of open
string background. 
At least for backgrounds related by exactly
marginal deformations,
this notion of manifest background independence can be made
precise using the language of connections in theory space \cite{RSZ,
SZ}, 
as we demonstrated in section \ref{ss31}.  
Starting from the tachyon vacuum
we can reach any D-brane configuration by simply considering the sliver 
associated with that particular BCFT. Note, however, that the
spirit here is somewhat different from that of boundary string field   
theory, $-$  in our approach
we always
represent the string field as an element of the state space of a
specific reference BCFT.   At the end it
should not matter which basis we use to expand the string field.
Indeed, we have seen in section \ref{s3} that physical quantities like 
ratios of tensions of D-branes do not depend on this choice.   

\item The structure of the string field theory action \refb{eo1}
is very similar in spirit to the action of $p$-adic string
theory~\cite{Brekke,0003278,0102071}. 
Both are non-local, and 
in both
cases the action expanded
around the tachyon vacuum is perfectly non-singular and has no
physical excitations. Yet in both cases the theory admits lump solutions
which support open string excitations. The D-$p$-brane solutions are
gaussian in the
case of $p$-adic string theory, and also in the case of vacuum string
field theory, 
although in this case the string field has
many components
corresponding to different oscillator excitations. It will be interesting
to explore if there is any deeper significance to this similarity. 

\item 
Vacuum string field theory
is much simpler than conventional cubic open string field theory.
Explicit analytic solutions of 
equations of motion are possible, as we saw
in this and the earlier papers. Also in this theory off-shell `tachyon'
amplitudes (and perhaps other amplitudes as well) around the tachyon
vacuum can be computed exactly up to overall
normalization. All this is commensurate with the fact that we have chosen
to expand the action around a simpler background. 
In this context we would
like to note that even the $p$-adic string action 
takes a simple form only
when expended about the tachyon vacuum. Once we introduce shifted fields
to expand the action around the D-brane background, the action looks
much more complicated.

\item 
Clearly the most
pressing problem at this stage is understanding the ghost sector. This
is needed not merely to complete the construction of the action, but
also  for
understanding gauge transformations in this theory. This, in turn is
needed  for classifying inequivalent classical solutions and the spectrum
of physical states around D-brane backgrounds. 
The knowledge of $\QQ$ will
also enable us to calculate quantum effects in this theory and determine
whether the theory contains in its full spectrum, the 
elusive closed string states. On this issue, 
at the end of section~\ref{ss31} we speculated 
whether integration of dimension 
$(1,1)$ closed string primaries over the bulk of the sliver could
yield a representation of closed string degrees of freedom in
the open string state space.

\end{itemize}

\medskip
In this paper we have seen that vacuum SFT incorporates
nicely the most attractive features of boundary 
SFT $-$ the automatic generation of correct tensions, 
and the description of solutions in terms of renormalization
group ideas. These features arise in vacuum string field
theory by taking into account the unusual geometrical definition
of the sliver state. The two present shortcomings of boundary SFT $-$
the tachyon vacuum being a singular endpoint
of the configuration space, and the difficulty of defining
the theory in the space of all backgrounds $-$ are avoided
in vacuum SFT. In a previous paper \cite{TWO},
we noted the remarkable simplicity of the algebraic approach
to the construction of D-brane configurations in vacuum SFT.  All in all,
we are led to believe that vacuum string field theory may provide a
surprisingly powerful and flexible approach to non-perturbative string
theory.

\medskip
\bigskip

\noindent{\bf Acknowledgements}:  
We would like to thank J. David,
D. Gaiotto, R. Gopakumar,  F. Larsen, 
J. Minahan, S. Minwalla, N. Moeller,
P. Mukhopadhyay, M.~Schnabl,
S. Shatashvili, S. Shenker, 
 A. Strominger,
W.~Taylor, E.~Verlinde and E. Witten for useful discussions.
The figures were prepared by Marty Stock, and we are
grateful to him for his careful work.
The work of L.R. was supported in
part by Princeton University
``Dicke Fellowship'' and by NSF grant 9802484.
The work of A.S. was supported in part by NSF grant PHY99-07949.
The work of  B.Z. was supported in part
by DOE contract \#DE-FC02-94ER40818.

\end{document}